\documentclass[9pt,twocolumn,twoside]{scrartcl}
\pdfoutput=1
\usepackage[utf8]{inputenc} 
\usepackage[T1]{fontenc}    
\usepackage{hyperref}       
\usepackage{url}            
\usepackage{booktabs}       
\usepackage{amsfonts}       
\usepackage{nicefrac}       
\usepackage{microtype}      
\usepackage{lipsum}
\usepackage{fancyhdr}       
\usepackage{graphicx}       
\usepackage{sidecap}
\usepackage{mdframed}
\usepackage{amsmath}
\usepackage[font=small,labelfont=bf, format = plain]{caption}
\usepackage[verbose=true,letterpaper]{geometry}
\AtBeginDocument{
  \newgeometry{
    textheight=9in,
    textwidth=18cm,
    top=1in,
    headheight=14pt,
    headsep=25pt,
    footskip=30pt
  }
}
\def\keywordname{{\bfseries \emph Keywords}}%
\def\keywords#1{\par\addvspace\medskipamount{\rightskip=0pt plus1cm
\def\and{\ifhmode\unskip\nobreak\fi\ $\cdot$
}\noindent\keywordname\enspace\ignorespaces#1\par}}

\graphicspath{{figures/}}     

\title{Glacial abrupt climate change as a multi-scale phenomenon resulting from monostable excitable dynamics}

\author{
  Keno Riechers \\
  Complexity Science \\
  Potsdam Institute for Climate Impact Research \\
  Potsdam, Germany\\
  \texttt{riechers@pik-potsdam.de} \\
  \and
  Georg Gottwald \\
  School of Mathematics and Statistics\\
  University of Sydney\\
  Sydney, Australia\\
  \and
  Niklas Boers \\
  Earth System Modelling - School of Engineering \& Design\\
  Technical University of Munich\\
  Munich, Germany
}

\begin{document}
\twocolumn[
\begin{@twocolumnfalse}
  \maketitle
  \thispagestyle{fancy}
  \begin{abstract}
    \noindent
    \textbf{Abstract.} Paleoclimate proxies reveal abrupt transitions of
    the North Atlantic climate during past glacial intervals known as
    Dansgaard-Oeschger (DO) events. A central feature of DO events is a
    sudden warming of about 10\(^{\circ}\)C in Greenland marking the
    beginning relatively mild phases termed interstadials. These exhibit
    gradual cooling over several hundred to a few thousand years until a
    final abrupt decline brings the temperatures back to cold stadial
    levels. As of now, the exact mechanism behind this millennial-scale
    variability remains inconclusive. Here, we propose an excitable model
    to explain Dansgaard-Oeschger cycles, where interstadials occur as
    noise-induced state space excursions. Our model comprises the mutual
    multi-scale interactions between four dynamical variables representing
    Arctic atmospheric temperatures, Nordic Seas’ temperatures and sea ice
    cover, and the Atlantic Meridional Overturning Circulation. The model’s
    atmosphere-ocean heat flux is moderated by the sea ice, which in turn
    is subject to large perturbations dynamically generated by fast
    evolving intermittent noise. If supercritical, perturbations trigger
    interstadial-like state space excursions during which all four model
    variables undergo qualitative changes that consistently resemble the
    signature of interstadials in corresponding proxy records. As a
    physical intermittent process generating the noise we propose
    convective events in the ocean or atmospheric blocking events. Our
    model accurately reproduces the DO cycle shape, return times and the
    dependence of the interstadial and stadial durations on the background
    conditions. In contrast to the prevailing understanding that DO
    variability is based on bistability in the underlying dynamics, we show
    that multi-scale, monostable excitable dynamics provides a promising
    alternative to explain millennial-scale climate variability associated
    with DO events. \keywords{Dansgaard--Oeschger Events $|$ Excitable
      Dynamics $|$ Multiscale System $|$ \(\alpha\)-stable Noise}
\end{abstract}
\vspace*{0.4cm}
\end{@twocolumnfalse}
]

\begin{mdframed}[linewidth=1pt, linecolor=black]
This Work has been accepted to Journal of Climate. The AMS does not guarantee that the copy provided here is an accurate copy of the Version of Record (VoR).
\end{mdframed}

\section{Introduction}
\begin{figure*}
  \centering
  \includegraphics[width = 0.9\textwidth]{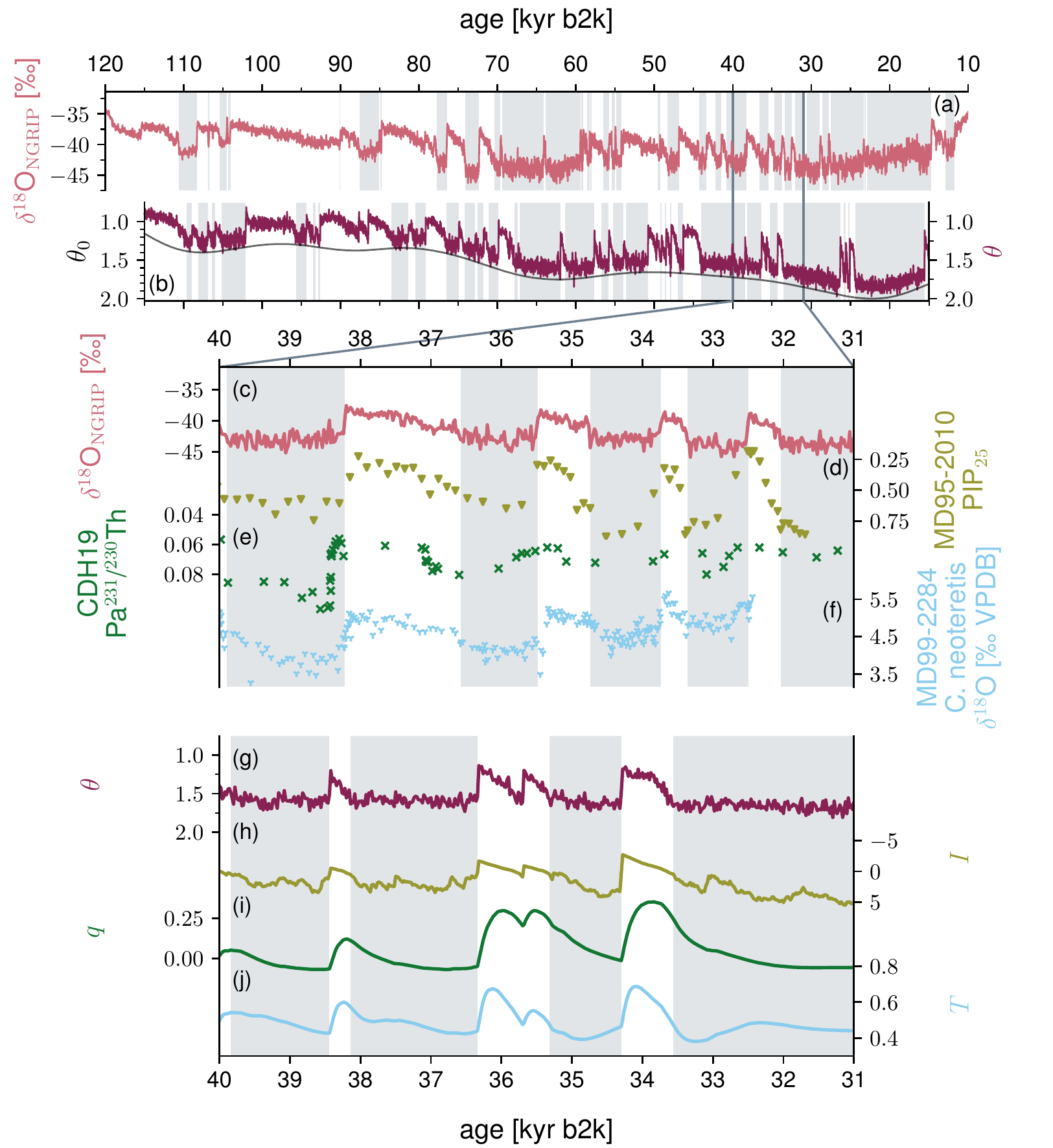}
  \caption{Paleoclimatic proxy evidence for characteristic features of DO
    variability together with corresponding results of our model defined by
    Eqs.~\ref{eq:1}--\ref{eq:5}. (a) 20-year mean NGRIP $\delta^{18}$O data
    \cite{Seierstad2014, Rasmussen2014a} interpreted as a qualitative
    proxy of air temperatures over Greenland (data available at:
    https://www.iceandclimate.nbi.ku.dk/data/, last access: 3 March
    2023). (b) Simulated Arctic atmospheric air temperatures $\theta$. (c)
    Zoom into the period 40--31~kyr~b2k of (a) --- the period was chosen
    due to the availability of proxy data. (d) PIP\(_{25}\) index from the
    marine sediment core MD95-2010 \cite{Sadatzki2020}. The PIP\(_{25}\)
    index is indicative of past sea ice cover at the core site with values
    of 1 and 0 corresponding to perennial sea ice and open ocean
    conditions, respectively. (e) Pa/Th ratios from the marine sediment
    core CDH19 from the Bermuda Rise as provided by \cite{Henry2016}. The
    ratios are interpreted as a direct measure of the AMOC strength, with
    lower values corresponding to stronger overturning and vice versa. (f)
    $\delta^{18}$O of the benthic species C. neoteretis \cite{Dokken2013}
    shown on a revised age-scale \cite{Berben2020, Sadatzki2020}.
    According to the most recent interpretation of the data, the benthic
    $\delta^{18}$O is mostly indicative of past intermediate to deep-ocean
    temperatures \cite{Ezat2014, Sadatzki2020} with higher values
    indicating colder temperatures and vice versa. Panels (g-j) show
    corresponding model results:
    (g) the Arctic atmospheric temperature \(\theta\), (h) the Nordic Seas'
    sea ice cover \(I\), (i) the AMOC strength \(q\) and (j) the Nordic
    Seas intermediate to deep water temperatures \(T\). The proxy records
    shown in (c)--(f) should be directly compared to the simulated
    trajectories shown in panels (g)--(j), respectively. All model
    variables are given in dimensionless units (see Appendix).}
\label{fig:01}
\end{figure*}

Stable water isotope records from Greenland ice cores provide evidence for
repeated abrupt climatic shifts during the last glacial interval.
Decadal-scale transitions from low to high values of $\delta^{18}$O
(Fig.~\ref{fig:01}a) indicate sudden warming events at the drilling site,
which are termed Dansgaard--Oeschger (DO) events \cite{Dansgaard1982,
  Dansgaard1984, Johnsen1992, Dansgaard1993, NGRIP2004}. The temperature
increases of approximately \(5-15^{\circ}\)C \cite{Jouzel1997, Johnsen2001,
  Landais2005, Huber2006a, Kindler2014} were followed by phases of milder,
yet moderately cooling temperatures called interstadials.
Typically, a final and more abrupt decline brought the climate back to a
colder state known as stadial climate. The millennial-scale successions of interstadials and stadials are often referred to as DO cycles. 

The signature of DO cycles is found in numerous paleoclimatic proxy records
around the globe, including speleothems and Antarctic ice cores
\cite{Voelker2002, Menviel2020}. These records show that the
Dansgaard-Oeschger cycles were most pronounced in the North Atlantic region
but were not limited to it. Instead they seized several components of the
global climate system. For instance, DO events are associated with
large-scale reorganizations of the Northern Hemisphere atmospheric
circulation
\cite{Ruth2007, Fischer2007,
  Schupbach2018}
including a northward shift of the ITCZ
\cite{Menviel2020} with strong impacts on the Asian and South American
Monsoon systems \cite{Wang2001, Kanner2012,
  Cheng2013,Zhang2017,Corrick2020}. Given the its strong local impact,
processes active in the North Atlantic region such as sea ice or deep water
formation are believed to be central to the triggering mechanism of DO
events \cite{Dokken2013, Boers2018, Vettoretti2018, Menviel2020,
  Sadatzki2020}.

To date there is no conclusive theory that fully explains the mechanism of
DO cycles. Several climatic components have been proposed to be relevant,
including ocean dynamics, atmospheric events, sea ice, ice sheets and
freshwater fluxes \cite{Broecker1985, Ganopolski2001,
  Petersen2013,Dokken2013, Zhang2014, Vettoretti2018, Boers2018, Li2019}.
Similarly, several dynamic mechanisms have been invoked to explain the DO
events, ranging from external drivers such as (periodically) changing
freshwater fluxes \cite{Ganopolski2001,Timmermann2003, Menviel2014} and
noise induced transitions between two stable states \cite{Ditlevsen1999,
  Ditlevsen2005, Ditlevsen2007, Lohmann2018b, Lohmann2022} to dynamically
self-generating mechanisms including self-sustained oscillations
\cite{Broecker1985, Broecker1990, Rasmussen2004, Dokken2013,
  Vettoretti2018, Boers2018, Vettoretti2022}.

Building on the results of \cite{Gottwald2021} we consider here a paradigm
for DO variability which to date has received only little attention. We
present an excitable multi-scale model for the dynamics of Greenlandic air
temperatures, Arctic sea ice, the Nordic Seas' intermediate to deep water
temperatures and the strength of the Atlantic meridional overturning
circulation. For certain parameters the model supports a single stable
fixed point corresponding to stadial climate conditions. In response to sea
ice perturbations above a critical threshold, the system takes prolonged
excursions in state space. Along this excitation path, it passes through a
region of slow transitive dynamics that can be identified with the
interstadial climate state consistently in all four model dimensions. The
required magnitude of the perturbations is generated by an intermittent
non-Gaussian driving noise detailed later. A DO-like excitation mechanism
for the North Atlantic glacial climate has previously been identified in an
intermediate complexity by \cite{Ganopolski2002}. Later,
\cite{Vettoretti2022} showed in a conceptual framework that shorter
interstadials could be explained as excitations. An important new aspect of
our model is that it effectively translates the strength of the exciting
noise pulse into the duration of the triggered interstadial therefore
providing a possible explanation for the great variety of the real-worlds
interstadials in terms of shape and duration.
Overall, the model reproduces the following five central aspects of
DO cycles in an interpretable manner (cf. Fig.~\ref{fig:01}):

\begin{enumerate}
\item \textbf{Shape of DO cycles:} The characteristic shape of DO cycles in
  the NGRIP \(\delta^{18}\)O record, as described above, is comprised of an
  abrupt warming followed by a gradual cooling and final stage of
  accelerated cooling back to stadial climate conditions. On close
  inspection there are many deviations from this archetypical shape such as
  short temperature declines within interstadials and vice versa, rebound
  events shortly before interstadial--stadial transitions or continuous
  interstadial--stadial cooling transitions without accelerated cooling
  (cf. Fig.~\ref{fig:01}a and c).

\item \textbf{Duration of stadials and interstadials:} Both, stadials and
  interstadials lasted from centuries to millennia. The ratio between
  interstadial and stadial duration, however, changed over time and was
  presumably influenced by external (orbital) forcing and internal forcing
  from the background climate state through the ice sheet configuration or
  atmospheric CO\(_{2}\) concentrations
  \cite{Rial2011, Roberts2017, Mitsui2017,Lohmann2018b, Boers2018,
    Zhang2021, Kuniyoshi2022, Vettoretti2022}. The early glacial
  (\(\sim\)115-71~kyr~b2k) was dominated by long lasting interstadials and
  relatively warm stadial
  conditions. The middle part of the glacial (\(\sim\)71-29~kyr~b2k) was
  characterized by frequent DO events with intermediate-length stadials and
  interstadials. Finally, colder stadials and very short interstadials
  prevailed during the later part of the last glacial
  (\(\sim\)29-14~kyr~b2k). Compare Fig.~\ref{fig:01}a and see
  \cite{Kindler2014} for temperature levels.



\item \textbf{In-phase sea ice dynamics:} Sea ice in the Nordic Seas and
  the North Atlantic varied in phase with Greenland temperatures. During
  stadials an extensive sea ice cover prevailed, whereas interstadials
  exhibited conditions ranging from open water to seasonal sea ice cover
  \cite{Li2005, Li2010, Dokken2013, Hoff2016, Sadatzki2020} (cf.
  Fig.~\ref{fig:01}d).
 
\item \textbf{Nordic Seas' temperature inversion:} Several studies report
  warming of the ice-covered stadial Nordic Seas at intermediate and large
  depth caused by a continued inflow of warm water masses from the south
  \cite{Rasmussen2004, Ezat2014, Sadatzki2020} (cf. Fig.~\ref{fig:01}f).
  Near-surface water temperatures show a similar pattern with an additional
  warming flush in synchrony with DO events \cite{Dokken2013}. The oceanic
  heat --- initially trapped under the sea ice --- is hypothesized to have
  abruptly warmed the polar atmosphere in response to sudden sea ice
  retreat during DO events \cite[e.g.]{Rial2011, Dokken2013, Boers2018,
    Vettoretti2018, Kuniyoshi2022}.

\item \textbf{AMOC switches:} Multiple lines of direct and indirect
  evidence, thoroughly summarized by \cite{Lynch-Stieglitz2017}, point to
  changes in the strength of the Atlantic Meridional Overturning
  Circulation (AMOC) in phase with Greenland temperatures, with weak (or
  no) overturning during (Heinrich) stadials and stronger overturning
  during interstadials (cf. Fig.~\ref{fig:01}e).
  An active AMOC is typically believed to have provided the necessary
  northward heat transport to explain the milder Arctic climate during
  interstadials.
\end{enumerate}

From a physical modelling point of view, the results presented below
suggest that the DO events may have been caused by complex multi-scale
interactions between several climate subsystems acting on separate time
scales: the ocean circulation, the sea ice, the large-scale atmosphere
ordered from slow to fast characteristic time scales, and intermittent
atmospheric or oceanic events on time scales faster than the sea ice time scale and comparable to the atmospheric time scale.


The paper is structured as follows:
We introduce the model in Section~\ref{sec:methods} and analyze its
dynamics in Section~\ref{sec:results}. We interpret the results in a
physical context and also perform a detailed model--data comparison in
terms of the above listed five key features. Section~\ref{sec:discussion}
discusses the results and relates them to previous research. We conclude in
Section~\ref{sec:conclusions} with a summary of our key findings.

\section{Methods}\label{sec:methods}
\subsection{Monostable excitable model of DO variability}

\begin{figure*}[t]
  \centering
  \includegraphics[width =0.8\textwidth]{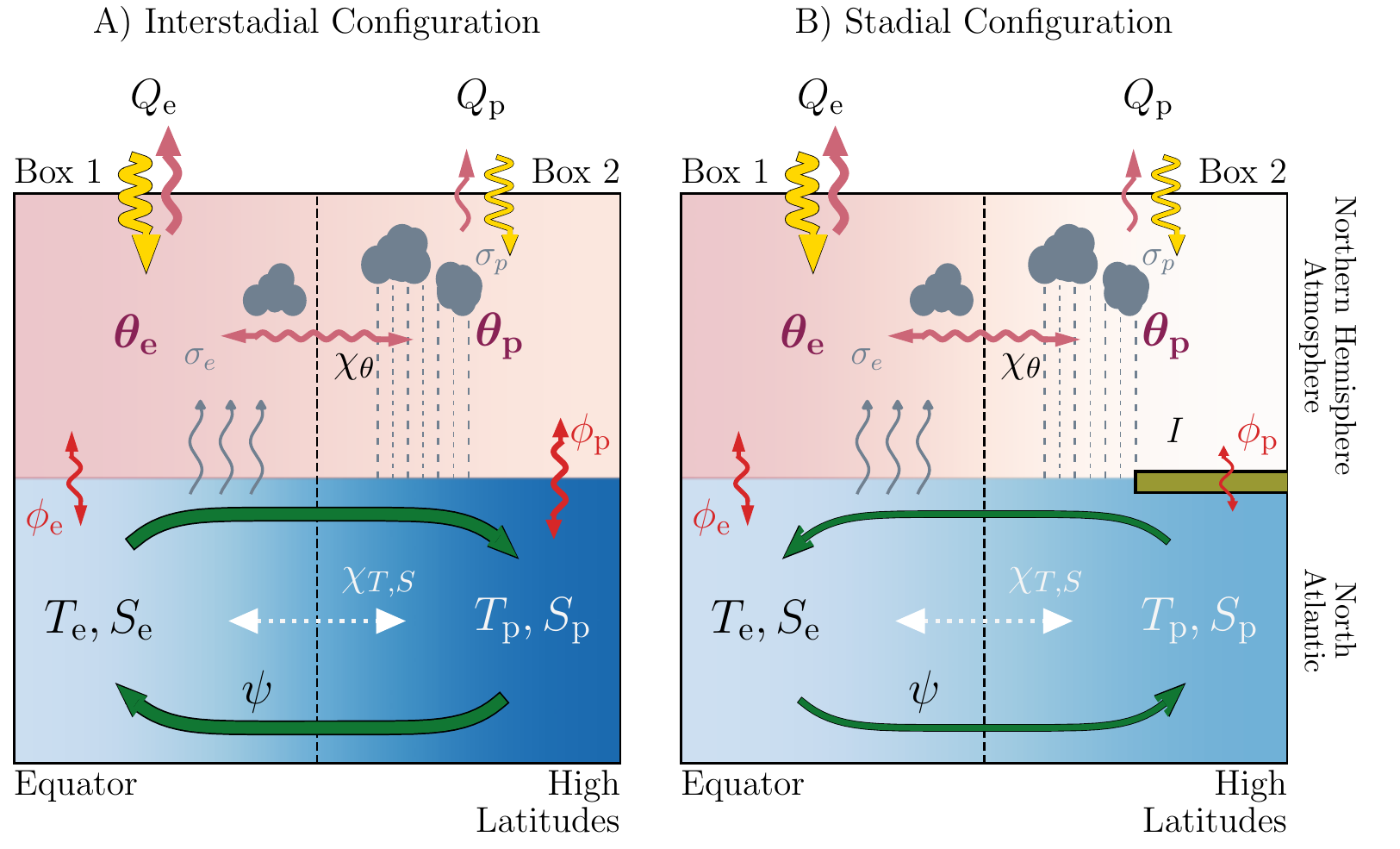}
  \caption{Schematic illustration of the excitable monostable model used to
    reproduce DO variability of the last glacial. The three key model
    components are the North Atlantic (bottom), the Northern Hemisphere
    atmosphere in the North Atlantic region (upper part) and the sea ice in
    the Nordic Seas (green bar on the right). Oceanic temperature and
    salinity in the equatorial and polar region are denoted as
    \(T_{\mathrm{e}}, S_{\mathrm{e}}\) and
    \(T_{\mathrm{p}}, S_{\mathrm{p}}\), respectively. The corresponding
    meridional gradients follow as \(T= T_{\mathrm{e}}-T_{\mathrm{p}}\) and
    \(S = S_{\mathrm{e}}-S_{\mathrm{p}} \). Analogously, the atmospheric
    temperature gradient
    \(\theta = \theta_{\mathrm{e}} -\theta_{\mathrm{p}}\) is given by the
    difference between equatorial and polar atmospheric temperatures. The
    solar differential heating caused by the difference in the net
    radiative heat fluxes \(Q_{\mathrm{e}}\) and \(Q_{\mathrm{p}}\)
    together with the atmospheric diffusive flux \(\chi_{\theta}\) sets the
    background hemispheric meridional temperature gradient \(\theta_{0}\).
    Atmosphere and ocean exchange heat in both the equatorial and the polar
    region (\(\phi_{\mathrm{e,p}}\)). The oceanic gradients \(T\) and \(S\)
    give rise to a meridional density gradient, that in turn drives the
    flow \(\psi\) which represents the AMOC. On top off that, the diffusive
    fluxes \(\chi_{T,S}\) counteract the existing gradients in the ocean.
    The oceanic salinity gradient \(S\) is maintained by constant fresh
    water flux in the equatorial and polar regions (not shown). The
    interstadial configuration (A) is characterized by the absence of sea
    ice in the high latitudes. This allows for an elevated heat flux
    \(\phi_{\mathrm{p}}\) resulting in a relatively cold ocean and a
    relatively warm atmosphere in the high latitudes. The pronounced
    meridional temperature gradient in the ocean yields a strong
    overturning flow \(\psi\). In the stadial configuration (B), the sea
    ice insulates ocean and atmosphere from one another. The
    sea-ice-covered high latitude ocean is relatively warm and the
    atmosphere above it is cold. The reduction of the oceanic temperature
    gradient causes a weak an reversed overturning flow. The derivation of
    the model equations under consideration of the illustrated flows is
    presented in the Appendix.}
  \label{fig:00}
\end{figure*}

From the slowest to the fastest time scale, the key components of our
conceptual multi-scale model are: (i) the Atlantic ocean represented by the
meridional temperature and salinity gradients $T(t)$ and $S(t)$ between the
Equator and the Nordic Seas, (ii) the Nordic Seas' sea ice extent $I(t)$,
(iii) the Northern Hemisphere atmosphere represented by the meridional
temperature gradient $\theta(t)$ and (iv) intermittent oceanic and
atmospheric anomalies $\xi_{t}$ and regular Gaussian atmospheric
fluctuations \(\zeta_{t}\). All variables are non-dimensionalized by a
suitable rescaling (see Appendix). The model set-up is schematically
illustrated in Figure~\ref{fig:00}. For the ease of notation, we will omit
the explicit time dependency of the four dynamical variables in the
following (i.e., we write \(\theta\) instead of \(\theta(t)\) for example);
fast driving noise processes will be subscripted with \(t\). A detailed
derivation of the individual model components is provided in the Appendix.
Here we give only a concise description of the model.

Following the classical Stommel model \cite{Stommel1961, Cessi1994}, on
the slowest time scale $\tau_{\mathrm{ocean}}$ the oceanic meridional
gradients of temperature and salinity evolve according to
\begin{align}
\label{eq:1}
\tau_{\mathrm{ocean}}\dot{T} &= -\gamma(I)(T-\theta) - (1 + \mu|T-S|)T,  \\
\label{eq:2}
\tau_{\mathrm{ocean}}\dot{S} &= \sigma - (1 + \mu|T-S|)S
\end{align}
and determine the strength of the density driven overturning flow
\begin{equation}
  \label{eq:AMOC}
  q = T - S, 
\end{equation}
which represents the AMOC strength. The ocean temperature $T$ is coupled to
the atmospheric temperature gradient $\theta$ via a heat flux whose
strength is moderated by the sea ice. This heat flux is effectively
represented by the mutual relaxation rate \(\gamma(I)\). The atmospheric
meridional temperature gradient $\theta$ relaxes against two opposing
temperature gradients, namely a prescribed background state \(\theta_{0}\)
which is determined by the net radiative heat fluxes and the atmospheric
diffusion and the oceanic gradient \(T\)
\begin{equation}
\label{eq:3}
\tau_{\mathrm{atm}} \dot{\theta} = -\eta ( \theta - \theta_{0}) - \gamma(I) (\theta-T)  + \zeta_t. 
\end{equation}
Here, $\tau_{\mathrm{atm}}<\tau_{\mathrm{ocean}}$ is a fast atmospheric
time scale and $\zeta_t$ denotes a white noise process that disturbs the
atmospheric dynamics. The ratio between the effective atmospheric
relaxation rate \(\eta\) and the mutual atmosphere--ocean relaxation rate
\(\gamma(I)\) determines the influences of \(\theta_{0}\) and \(T\) on
\(\theta\).

It is widely accepted that changing background climate conditions strongly
influenced DO variability over the course of the last glacial
\cite{Rial2011, Roberts2017, Mitsui2017, Boers2018}. This effect is
considered in our model by altering the atmospheric background state
\(\theta_{0}\) over time according to
\begin{equation}
\label{eq:6}
\theta_{0}(t) = 1.59 + 0.23 \delta^{18}\mathrm{O}^{\ast}_{\mathrm{LR04}}(t).
\end{equation}
The normalized benthic stable isotope data
\(\delta^{18}\mathrm{O}^{\ast}_{\mathrm{LR04}}(t) \) is indicative of past
global ice volume changes \cite{Lisiecki2005}. We argue that a colder
background climate increases the atmospheric background temperature
gradient \(\theta_{0}\) due to Arctic amplification
\cite{Masson-Delmotte2006}.

On an intermediate time scale $\tau_{\mathrm{ice}}$ with
$\tau_{\mathrm{ocean}} > \tau_{\mathrm{ice}}>\tau_{\mathrm{atm}}$, the
temporal evolution of the sea ice is given by the seasonally averaged
Eisenman model \cite{Eisenman2012, Lohmann2021a}
\begin{equation}
\label{eq:4}
\tau_{\mathrm{ice}}\dot{I} = \Delta \tanh\left(\frac{I}{h}\right) - R_{0}H(I)I - L_{0} + L_{1}\theta - L_{2}I + \xi_t,
\end{equation}
where $H(I)$ denotes the Heaviside function and the term \(L_{1}\theta\)
represents the influence of the atmosphere on the sea ice formation, with
large atmospheric temperature gradients --- i.e. colder temperatures at
high northern latitudes --- fostering sea ice growth and vice versa. We
assume that the ocean's surface layer which is not resolved in our box
model, is dominated by atmospheric temperatures. This justifies the direct
coupling of the sea ice to the atmosphere. The remaining terms represent
the ice--albedo feedback, sea ice export and the net incoming and outgoing
radiation linearized with respect to \(I\), respectively. For a detailed
discussion of the sea ice model, please see Appendix. In the original model
formulation, the non-dimensionalized variable \(I\) represents the sea ice
thickness over a horizontally homogeneous ocean column
\cite{Eisenman2012}. Therein, values of \(I<0\) correspond to an ice free
state. Here, we interpret \(I\) as a stylized representation of the
annually averaged sea ice volume in the Nordic Seas. The time scale
\(\tau_{\mathrm{ice}} = 200\;\mathrm{years}\) may be understood as the rate
at which the multi-year sea ice front advances southwards. The stochastic
process $\xi_t$ models fast intermittent random sea ice retreat events that
will be explained in greater detail below.

The sea ice itself couples back to the ocean and atmosphere dynamics by
acting as a dynamic insulator \cite{Dokken2013, Boers2018} and modifying
the respective mutual relaxation rate according to
\begin{equation}
\label{eq:5}
\gamma(I) = \gamma_{0} +  \frac{\Delta \gamma}{2}\left[\tanh\left(\frac{-(I-I_{0})}{\omega}\right)+1\right].
\end{equation}
In the presence of a stadial sea ice cover (\(I>0.5\)) the polar ocean is
shielded from the atmosphere and the mutual heat flux is heavily
suppressed. In contrast, a reduced interstadial sea ice cover yields a
strong atmosphere--ocean heat flux and correspondingly a high mutual
relaxation rate \(\gamma\). The choice of \(I_{0}\) determines what should be regarded as intermediate sea ice cover in our model.

The next paragraph establishes a correspondence between the model variables
and the considered climate variables, or more precisely their respective
proxy variables. Although $T$ and \(\theta\) are gradients, we may compare
them directly with observations for oceanic and atmospheric temperature
proxies from high northern latitudes, respectively, with larger gradients implying colder temperatures in the polar region. This is justified since the
comparably much larger size of the equatorial region compared to the polar
region implies substantially larger heat capacities for the considered
equatorial boxes. We thus ascribe changes in the gradients mostly to
changes in the polar regions and
interpret \(T\) and \(\theta\) as direct counterparts for intermediate and
deep ocean temperature proxy records from the Nordic Seas (compare
Fig.~\ref{fig:01}f with \ref{fig:01}j; increases in the gradients reflect
cooling of the polar boxes) and $\delta^{18}$O records from Greenland ice
cores (compare Fig.~\ref{fig:01}c with \ref{fig:01}h), respectively. The
comparison of \(q\) with proxies for past AMOC strength is straightforward
(compare Fig.~\ref{fig:01}e with \ref{fig:01}j) and so is the comparison of
\(I\) with proxies for past sea ice extent (compare Fig.~\ref{fig:01}d with
\ref{fig:01}i).

Values of the parameters used in our numerical simulation are summarized in
Table~\ref{tab:01}.

\begin{table}[h]
  \centering
  \begin{tabular}{|l p{5cm} l|}
    \hline 
      Parameter & Definition & Value \\
    \hline
    \multicolumn{3}{|l|}{\textbf{time scales}}\\
    \hline
    \(\tau_{\mathrm{ocean}}\) & oceanic time scale in years & 800  \\
    \(\tau_{\mathrm{ice}}\) & sea ice time scale in years & 200 \\
    \(\tau_{\mathrm{atm}}\) & atmospheric time scale in years  & 0.6 \\
    \hline
    \multicolumn{3}{|l|}{\textbf{atmosphere--ocean model}}\\
    \hline
    \(\gamma_{0}\) & atmosphere--ocean relaxation rate at full sea ice cover&0.5 \\
    \(\Delta \gamma\) & amplitude of the sea ice's insulation effect & 3.5\\
    \(\eta\) & atmospheric heat dissipation rate  & 4 \\
    \(\mu\) & flux parameter & 7.5 \\
    \(\sigma\) & freshwater influx & 0.7 \\
    \hline
    \multicolumn{3}{|l|}{\textbf{sea ice model}}\\
    \hline
    \(L_{0}\)& 0th order sea ice OLR & 1.75 \\
    \(L_{1}\)& linear dependence of sea ice OLR on the atmosphere & 1.85 \\
    \(L_{2}\)&  linear dependence of sea ice OLR on the sea ice & 0.35 \\
    \(\Delta\)& strength of sea ice albedo feedback & 0.25\\
    \(h\)& characteristic sea ice albedo feedback scale & 0.08 \\
    \(R_{0}\)& rate of sea ice export & 0.4 \\
    \(\omega\) & characteristic insulation scale of sea ice & 0.8\\
    \(I_{0}\) & sea ice value, at which half of the insulation effect is reached & -0.5\\
    \hline 
  \end{tabular}
  \caption{Model parameters used in all simulations, unless stated
    otherwise. The parameters are chosen to reproduce the key features of
    DO events.}
  \label{tab:01}
\end{table}


For the parameter configuration specified in Tab.~\ref{tab:01} the
deterministic model defined by Eqs.~\ref{eq:1}--\ref{eq:3} and
Eqs.~\ref{eq:4}--\ref{eq:5} yields monostable dynamics for
\(\theta_{0}>1.275\). 
However, it also features a region of slow transitive dynamics located
where the nullclines of the atmosphere and sea ice variables are closest
(c.f. Fig.~\ref{fig:deterministic_trajectories}a and f). We will show later
that this meta-stable state in the model's state space can be identified
with interstadial climate conditions of the North Atlantic region. In order
to make this meta-stable state accessible to the dynamics, in the
following, we introduce the noise processes \(\xi_{t}\) and \(\zeta_{t}\)
that mimic the effect of unresolved events occurring on time scales faster
than the characteristic time scales $\tau_{\mathrm{ice}}$ and
$\tau_{\mathrm{atm}}$ of the sea ice and atmosphere dynamics, respectively.
Notice that both, the monostability and the excitability depend on the
specific choice of parameters. For other values, the dynamical features of
the model may differ.

\subsection{Stochastic (intermittent) forcing processes
  \(\boldsymbol{\xi_{t}}\) and \(\boldsymbol{\zeta_{t}}\)}
\label{sec:noise}

\begin{figure}[h]
  \centering
  \includegraphics{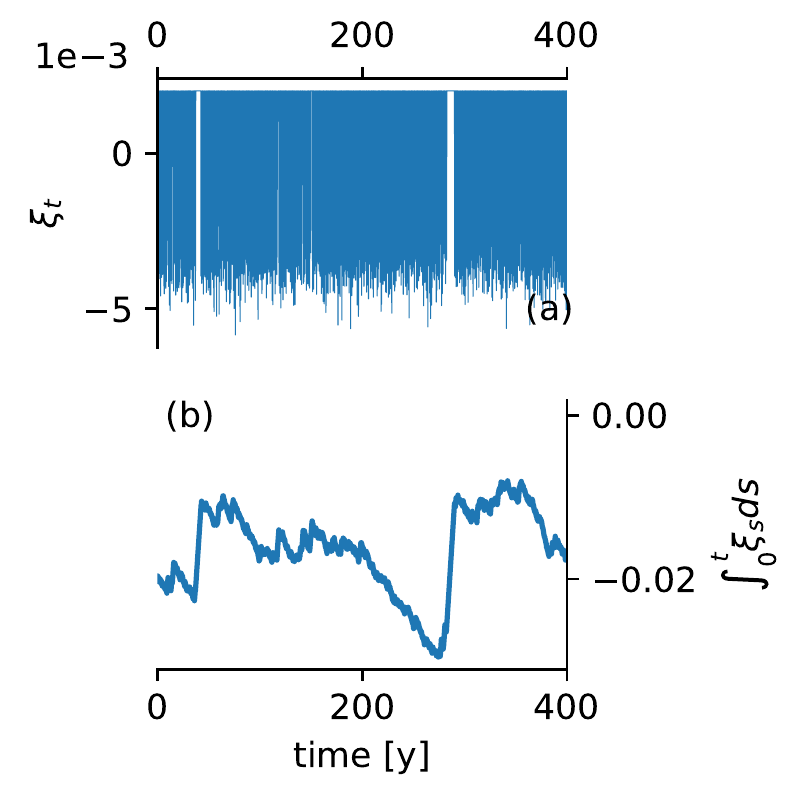}
  \caption{Illustration of the driving noise process $\xi_{t}$ acting on
    the sea ice (a) and its integrated form $\int_{0}^{t}\xi_{s} ds$ (b).
    It is clearly seen how prolonged laminar phases of the driving noise
    $\xi_t$ result in jumps in the integrated $\int_{0}^{t}\xi_{s} ds$.
    These jumps may correspond to supercritical sea ice removals which in
    turn trigger DO events in our excitable model.}
  \label{fig:noise}
\end{figure}

\begin{table}[h]
  \centering
  \begin{tabular}{|l p{5cm} c|}
    \hline
    Parameter & Definition & Value \\
    \hline
    \multicolumn{3}{|l|}{\textbf{atmospheric noise $\zeta_t$}}\\
    \hline
    \(\sigma_{\theta}\) & amplitude of the atmospheric noise & 0.04\\
    \hline
    \multicolumn{3}{|l|}{\textbf{sea ice noise $\xi_t$ during stadials}}\\
    \hline
    c  &laminar forcing strength & 0.2 \\
    \(\sigma_{\mathrm{tur}}\) & amplitude of the Brownian motion during turbulent phase & 0.01 \\
    \(k = 1/\alpha \) &shape parameter of Pareto distribution & 0.62\\
    \(\sigma_{\mathrm{lam}} \) &scale parameter of the Pareto distribution& 2\\
    \(\mu_{\mathrm{lam}} = \sigma_{\mathrm{lam}}/k\)& location parameter of the Pareto distribution & 3.2\\
    \hline
    \multicolumn{3}{|l|}{\textbf{sea ice noise $\xi_t$ during interstadials}}\\
    \hline
    \(\sigma_{I}\) &amplitude of interstadial Brownian motion & 0.006\\
    \hline
  \end{tabular}
  \caption{Parameters for the stochastic processes $\zeta_t$ and $\xi_t$ which drive the atmosphere and sea ice in \ref{eq:3} and \ref{eq:4}, respectively.} 
  \label{table:noise}
\end{table}

The atmosphere variable \(\theta\) is assumed to be subjected to Gaussian
white noise forcing $\zeta_t = \sigma_{\theta} \dot W$ with standard
Brownian motion $W$. This noise can be motivated as the effective stochastic
effect of unresolved strongly chaotic atmospheric fluctuations on $\theta$
\cite{Hasselmann1976}. It does not substantially affect the
dynamics of Eqs.~\ref{eq:1}--\ref{eq:3} and
Eqs.~\ref{eq:4}--\ref{eq:5} but generates more realistic fluctuations
of \(\theta\) in accordance with the NGRIP $\delta^{18}$O record.

The sea ice noise $\xi_t$, however, plays a crucial role in triggering DO
events. To generate rare but large forcing events which allow the system to
leave the stadial fixed point and enter the meta-stable state, non-Gaussian
noise is required. \cite{Gottwald2021} showed that $\alpha$-stable noise,
which is characterized by the occurrence of discrete jumps, can be
dynamically generated in a multi-scale setting to produce abrupt warming
events in a Stommel model driven by a simple sea-ice model. Here, we
postulate that sea ice is subjected to rare and intermittent fast
processes. According to the theory laid out in
\cite{Gottwald2013,Gottwald2013a}, \cite{Gottwald2017} and
\cite{Gottwald2021}, this forcing, when integrated, gives rise to an
effective $\alpha$-stable component in the resulting dynamics of the sea
ice. The discrete jumps of the generated \(\alpha\)-stable process
represent large stochastic sea ice melting events.

By controlling the mutual relaxation rate \(\gamma(I)\) the sea ice in turn
drives the atmospheric and oceanic variables \(\theta\) and \(T\) with
emergent non-Gaussian noise. Indeed, signatures of non-Gaussian
$\alpha$-stable noise have been detected by \cite{Ditlevsen1999} in the
calcium concentration record of the GRIP ice core \cite{Fuhrer1993}.

We propose two possible physical mechanisms which may constitute such
intermittent forcing on the sea ice: oceanic convective events and
atmospheric anomalies. During stadials, the Nordic Seas' sea ice is
shielded from the warmer subsurface and deep waters by a thin layer of cold
and fresh water \cite{Dokken2013, Sadatzki2020}. We hypothesize that
intermittent convective events may temporarily remove this layer and melt
sea ice from below, efficiently opening up polynya through which oceanic
heat could be released to the atmosphere \cite{Vettoretti2018}. Either,
after locally releasing sufficient heat, a stable stratification of the
ocean reestablishes and the polynya refreeze. Or, the convective events
might remove a critical amount of sea ice and push the system into the
meta-stable interstadial state.

Strong atmospheric anomalies constitute another possible source of
intermittent sea ice forcing. \cite{Kleppin2015} and \cite{Drijfhout2013}
describe --- although in a different setting --- how persistent
atmospheric anomalies can drive the high northern latitude climate into a
substantially altered state. Storms or baroclinic instabilities could also act as initiators of oceanic convective events.  

We postulate that the above described mechanisms giving rise to
intermittent anomalous forcing events are only active during stadials.
During interstadials, we argue that convectively driven sea ice removal
should not have a strong impact on the already northward displaced sea ice
edge. Travelling northward the warm Atlantic inflow looses too much heat
before it can be subducted under the sea ice. Thus, heat cannot efficiently
accumulate underneath the sea ice which is a precondition for large
convective sea ice removal. Similarly, we argue that atmospheric anomalous
forcing events require a certain meridional temperature gradient and a
stadial configuration of the jet stream \cite{Li2019}. Therefore, we
impose a Gaussian white noise forcing of the sea ice dynamics with
$\xi_t = \sigma_{I} \dot W$ with standard Brownian motion $W$ if no
pronounced stadial sea ice cover is present ($I<0.5$). Since sea ice
fluctuations are smaller, the smaller the sea ice extent, we choose
relatively small \(\sigma_{I}\) as compared to the sea ice fluctuation
emerging from the more complex stadial driving noise which we describe in
the following.

To model intermittent convective events or atmospheric anomalies, we follow
\cite{Gottwald2013,Gottwald2013a} and design a (mean-zero) process which
consists of a succession of turbulent and laminar periods. 
The forcing in the laminar periods is set to a constant $\xi_t = -c$ whereas during the turbulent periods it fluctuates around $\xi_t = c$ according to standard Brownian motion with $\xi_t = c + \sigma_{\mathrm{tur}} \dot{W}$. 
The respective durations of these phases are themselves random variables.
In particular, durations of the laminar period \(\tau_{\mathrm{lam}}\) can be
arbitrarily long and are distributed according to a Pareto law
\begin{equation}
  \label{eq:gprnd}
\tau_{\mathrm{lam}} \sim \frac{1}{\sigma_{\mathrm{lam}}}\left(1 + k \left[\frac{\tau_{\mathrm{lam}} - \mu_{\mathrm{lam}}}{\sigma_{\mathrm{lam}}}\right]\right)^{(-1/k + 1)}
\end{equation}
with shape parameter $k=1/\alpha$, scale parameter $\sigma_{\mathrm{lam}}$,
and location parameter \(\mu_{\mathrm{lam}} = \sigma_{\mathrm{lam}}/ k\).
Hence, laminar periods last on average for
$\mathbb{E}[\tau_{\mathrm{lam}}] = \sigma_{\mathrm{lam}}\alpha^{2}/(\alpha-1)$
time units (assuming $\alpha>1$). The durations of turbulent periods
\(\tau_{\mathrm{tur}}\) are uniformly distributed around the mean
$\bar{\tau}_{\mathrm{tur}} =\mathbb{E}[\tau_{\mathrm{lam}}] $ with
\( \tau_{\mathrm{tur}} \sim \bar{\tau}_{\mathrm{tur}} + U[-\bar{\tau}_{\mathrm{tur}}/2 , +\bar{\tau}_{\mathrm{tur}}/2]\), where
\(U[a,b]\) denotes the uniform distribution between the limits \(a\) and
\(b\). When such a process is integrated, during the laminar periods we
obtain ballistic flights with $\int_{0}^{t} \xi_{s} \mathrm{d}s =-c t $. The heavy
tail of the Pareto distribution assigns a probability of
$(\alpha\sigma_{\mathrm{lam}})^{\alpha}\tau^{-\alpha}$ to durations
$\tau_{\mathrm{lam}} > \tau$, and allows for $\alpha<2$ for non-vanishing
probabilities of ballistic flights of arbitrary lengths. This renders
\(\int\xi_{t}dt\) an effective $\alpha$-stable process. This mechanism of
intermittent laminar dynamics generating $\alpha$-stable noise is
illustrated in Figure~\ref{fig:noise}, where we show the stochastic process
$\xi_t$ and its integrated form.

Long lasting laminar forcing events may remove large amounts of sea ice and
thus entail an abrupt shift in the atmosphere--ocean heat flux which
determines the climatic state of the coupled atmosphere-ocean model (cf.
Fig.~\ref{fig:deterministic_trajectories}). Such dynamically generated
perturbations are capable of inducing meta-stable interstadial dynamics in
our model as we will show in the next section. In Table~\ref{table:noise}
we list the parameters used to generate the noise.


\section{Results}
\label{sec:results}

We first analyze the response of the deterministic coupled model to imposed
perturbations of the sea ice cover with $\zeta_t=\xi_t=0$ and a constant
climate background temperature $\theta_0$ in
Sect.~\ref{sec:results}\ref{sec:deterministic_dynamics}. We show that long
lasting interstadial-like excursions occur as a consequence of
supercritical sea ice perturbations of the stable stadial state. This
phenomenon results from a complex interplay of the three separate time
scales and the slow deterministic dynamics in a particular meta-stable
region of the state space, which is characterized by high proximity of the
\(\theta\) and \(I\) nullclines (cf.
Fig.~\ref{fig:deterministic_trajectories}).

In Sect.~\ref{sec:results}\ref{sec:noise_results} we show that the intermittent
noise $\xi_t$ is capable of generating such supercritical perturbations to
the sea ice acting as triggers of interstadials. Finally, the full
stochastic model, coupled to the background climate, is run over the entire
last glacial in Sect.~\ref{sec:results}\ref{sec:transient}.

Notice that in this section, we use the term \textit{nullcline} in slight deviation from its formal definition. Formally, the \textit{nullcline} of any of the model's variables is the set of points in the four dimensional state space where its derivative is zero. Let \(\Omega = (\theta, T, q)\) denote the state of the atmosphere--ocean model component of the coupled model. Conditioned on a given value for the sea ice, the atmosphere--ocean model has either one or three fixed points in the subspace spanned by \(\theta\), \(T\) and \(q\). For ease of notation, we refer to the set \(\{(\theta, I):\mathrm{d}\Omega /\mathrm{d}t = 0\})\) as the \(\theta\)-nullcline in the \(\theta-I\) plane (\(T\)- and \(q\)-nullclines are defined correspondingly). Since the temporal evolution of sea ice only depends on \(\theta\), the \(I\)-nullcline shall be defined as the set of points where \(\mathrm{d}I /\mathrm{d}t = 0\) in the \(\theta-I\) plane.


\subsection{Deterministic response to sea ice perturbations}
\label{sec:deterministic_dynamics}
\begin{figure*}
  \centering
  \includegraphics[width = 0.9\textwidth]{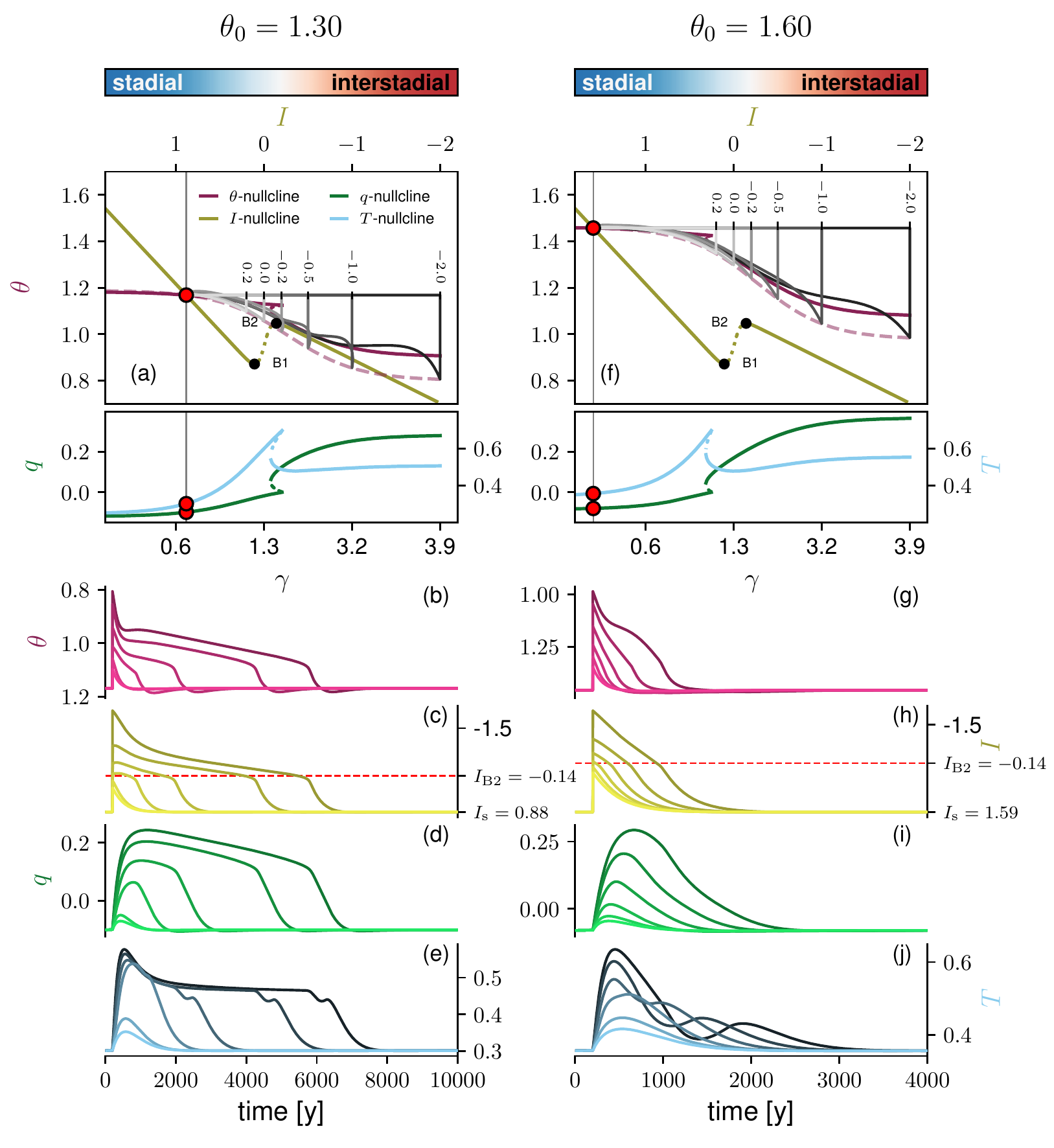}
  \caption{Trajectories of the deterministic system defined
    by Eqs. \ref{eq:1}--\ref{eq:5} initialized in the stable state and
    subjected to instantaneous sea ice retreat at \(t = 200\) with
    \(\zeta_{t} = \xi_{t} =0\). Six different sea ice perturbations have
    been applied: \(I_{\mathrm{p}} = \{0.2, 0.0, -0.2, -0.5, -1.0, -2.0\}\)
    (gray horizontal lines in a and f). Panels (a-e) and panels (f-j)
    correspond to warmer (\(\theta_{0} = 1.3\)) and colder
    (\(\theta_0 = 1.6\)) background climate, respectively. Panels (a) and
    (f) show the trajectories in the \(\theta\)--\(I\) plane together with
    the corresponding nullclines of all four model variables. A prescribed
    value for the sea ice variable (top axis) determines the
    atmosphere--ocean relaxation rate \(\gamma(I)\) (bottom axis), which in
    turn sets the stable fixed points for the coupled atmosphere--ocean
    model comprised of \(\theta, T\) and \(q\). Strong sea ice cover and a
    low mutual relaxation rate yield a cold polar atmosphere (large
    \(\theta\)), warm intermediate and deep waters in the Nordic Seas
    (small \(T\)) and a weak AMOC (small \(|q|\)). This configuration
    corresponds to stadial climate conditions as inferred from proxy
    records. A small sea ice cover reversely entails a warm polar
    atmosphere (small \(\theta\)), cold Nordic Seas (large \(T\)) and an
    active AMOC (large \(|q|\)) which can be identified with interstadial
    climate conditions. Since only \(\theta\) couples back to the sea ice
    \(I\), intersections of the \(I\)- and \(\theta\)-nullcline constitute
    fixed points of the entire system in the \(\theta-I\) plane with values
    for \(T\) and \(q\) (red dots) following from the heat exchange rate
    associated with this intersection. The remaining panels show the
    trajectories of the individual system variables against time for the
    different sea ice perturbations with darker colors corresponding to
    stronger sea ice removal. The horizontal dashed lines in panels (c) and
    (h) mark the critical sea ice threshold \(I_{B2}\) which constitutes
    the highest possible sea ice cover in the low-ice regime. Similarly,
    the sea ice bifurcation point \(I_{B1}\) marks the lowest possible sea
    ice cover in an ice-rich regime viewed in the \(\theta\)--\(I\) plane.}
\label{fig:deterministic_trajectories}
\end{figure*}
To understand how our model may explain DO variability, we investigate how
the deterministic system given by Eqs.~\ref{eq:1}--\ref{eq:3} and
Eqs.~\ref{eq:4}--\ref{eq:5} with \(\zeta_t = 0, \xi_t = 0\) recovers after
large imposed perturbations of the sea ice variable. We consider two
different choices of \(\theta_{0}= 1.3\) and \(\theta_{0} = 1.6\),
corresponding to warm and intermediate glacial climate backgrounds,
respectively (Fig.~\ref{fig:deterministic_trajectories}).
In simulations initialized in the stadial stable state, at time
\(t_{\text{p}} = 200\;\mathrm{years}\), the sea ice is abruptly removed by
manually setting \(I(t) = I_{\text{p}}\) with
\(I_{\text{p}} = \{0.2, 0, -0.2, -0.5, -1, -2\}\). Subsequently we let the
system evolve freely and relax back to the unique stadial fixed point. The
character of the relaxation depends on the strength of the perturbation and
on the value of the climate background $\theta_0$. In general, in the
warmer background climate with \(\theta_{0}\), the sea-ice nullcline and
the atmosphere nullcline are much closer together (cf.
Fig.~\ref{fig:deterministic_trajectories}). This allows for pronounced
interstadials, resembling real-world interstadials from the early glacial.
For intermediate background climate (\(\theta_{0}\)) the distance between
the two nullclines is larger. This yields shorter interstadials similar to
those observed during the mid-glacial. We observe the following distinct
phases in the system response.

\textbf{Phase A: immediate atmospheric response.} 
Over short time intervals the slow ocean dynamics $T$ and $S$ can be
considered as constants for the faster atmosphere dynamics $\theta$. Hence,
the atmospheric temperature gradient rapidly
decreases (i.e., the Arctic temperature increases) to the value $\theta_{\mathrm{p}}$
(see Fig.~\ref{fig:deterministic_trajectories}b,g), which is approximately
given as the solution to
$\gamma(I_{\mathrm{p}})(\theta_{\mathrm{p}}-T_{\mathrm{s}}) +
\eta(\theta_{\mathrm{p}}-\theta_{0})=0$ (shown as a light dashed line in
Fig.~\ref{fig:deterministic_trajectories}a and f as a guide to the eye).
The subscript \({\mathrm{s}}\) denotes the respective stadial fixed point
values.
Physically, $\theta_{\mathrm{p}}$ is determined by a new balance between
the competing influences of \(\theta_{0}\) and of \(T_{\mathrm{s}}\). Suddenly exposed to the warm stadial ocean (small \(T\)) which was
previously shielded from the atmosphere by the stadial sea ice cover, the high latitude atmosphere now takes up much more oceanic heat. 
The slow model time scale \(\tau_{\mathrm{ocean}}\) may in this context be
interpreted as the large heat capacity of the Nordic Seas, allowing heat
release to the atmosphere while their own temperature \(T\) remains
unchanged on fast time scales.
%


\textbf{Phase B: system-wide relaxation.} We now discuss how the system
relaxes form the perturbed state
\((\theta_{\mathrm{p}}, T_{\mathrm{s}}, q_{\mathrm{s}}, I_{\mathrm{p}})\)
back to the unique stable fixed point
\((\theta_{\mathrm{s}}, T_{\mathrm{s}}, q_{\mathrm{s}}, I_{\mathrm{s}})\).
Notice that the sea ice bifurcation point \(I_{\mathrm{B}2}\) marks the
lowest sea ice cover of the sea ice nullcline's stable branch with reduced
sea ice cover (cf. Fig.~\ref{fig:deterministic_trajectories}). The
relaxation is qualitatively different depending on whether the perturbation
brings the sea ice
to the low-ice regime (\(I_{\mathrm{p}}< I_{\mathrm{B}2}\)) or not. If that
is the case, the system takes a prolonged excursion in state space with a
two-stage relaxation process, of which the first stage can be identified
with interstadial climate conditions. We call this scenario, which involves
responses of all model variables, the supercritical case. If
\(I_{\mathrm{p}}> I_{\mathrm{B}2}\), then the system shows a straight
relaxation back to stadial conditions without any substantial response of
the oceanic variables to the initial perturbation. We term this scenario
the subcritical case. The existence of both subcritical perturbations which
rapidly relax back to the steady state and supercritical perturbations
which cause long transitory dynamics back towards the steady state
involving several time scales is a hallmark of so called excitable media
often found in neurophysiological systems \cite{FitzHugh1961, Nagumo1962}.

\textbf{Phase B: subcritical case
  \(\boldsymbol{I_{\mathrm{p}}> I_{\mathrm{B}2}}\).} In the subcritical
case the system remains in an ice-rich state which due to the albedo
feedback facilitates a fast regrowth of the sea ice.
Consequently, the dynamics of $I$ and \(\theta\) jointly relax back
straight to the stadial equilibrium as the regrowing sea ice increasingly
shields the atmosphere from the warmer ocean. The oceanic variables on the
other hand show hardly any response due to their higher inertia and the
fast sea ice recovery. Qualitatively the general system response to
subcritical perturbations is the same for \(\theta_{0} = 1.3\) and for
\(\theta_{0} =1.6\).

\textbf{Phase B: supercritical case
  \(\boldsymbol{I_{\mathrm{p}}< I_{\mathrm{B}2}}\) - Stage 1.} In contrast,
in the supercritical case a phase of slowed-downed sea ice recovery occurs,
giving the oceanic variables enough time to respond to the perturbation
such that all model variables temporarily assume interstadial
configurations, i.e. in addition to the elevated Arctic atmospheric
temperature (low gradient \(\theta\)) and the reduced Nordic Seas' ice cover (low
\(I\)), the AMOC assumes its strong circulation mode (large \(q\)) and the
Nordic Seas cool at intermediate and large depth (high gradient \(T\)) in agreement
with proxy evidence (cf. Fig.~\ref{fig:01}). 

As a consequence of the substantially reduced sea ice cover the ice--albedo
feedback now inhibits the reformation of the sea ice and in the presence of
warm atmospheric conditions \(\theta_{\mathrm{p}}\) the sea ice regrows at
a slow rate towards \(I_{\mathrm{B}2}\) or retreats even further if the
perturbed state is located left of the sea ice nullcline in the
\(\theta\)-\(I\) plane (cf. Fig.~\ref{fig:deterministic_trajectories}a). As
the atmosphere continuously dissipates the heat it receives from the ocean
while relaxing towards \(\theta_{0}\), the ocean starts to notably cool
(increase in \(T\)). In turn, the oceanic cooling reduces the atmospheric
uptake of oceanic heat and as a result the Arctic atmosphere simultaneously cools (see the pronounced increase in \(\theta\) in
Fig.~\ref{fig:deterministic_trajectories}b and g shortly after the perturbations).
This effect inevitably prevents a stabilization of the system in the
low-ice regime and ensures that the sea ice eventually regrows. A second
consequence of the Nordic Seas' cooling is the transition of the AMOC from
a salinity-driven weak mode to the temperature-driven strong mode. This
completes the interstadial configuration of the four model variables and
allows us to identify this first stage of the supercritical relaxation with
the interstadial climate. Fully reactivated, the AMOC's northward oceanic
heat transport stabilizes the temperatures of the Nordic Seas and over the
course of the reaming interstadial the slow adjustments of the other
variables are driven by the incremental sea ice regrowth.



The nature and duration of the interstadial stage depends on both the size
of the initial sea ice perturbation $I_{\mathrm{p}}$ and the closeness of
the sea ice nullcline and the atmosphere nullcline. If the two nullclines
are close (as for $\theta_0=1.3$) the dynamics of $\theta$ and $I$ is
comparably slow (cf. the dynamics for $\theta_0=1.3$ in
Fig.~\ref{fig:deterministic_trajectories}). Sufficiently strong sea ice
perturbations then result in pronounced interstadials during which all
system variables hardly change over an extended period of time. If the
nullclines are further apart (as for $\theta_0=1.60$) the interstadial
state is less pronounced and characterized by a gradual change in all model
variables, driven by steady sea ice regrowth. The separation of the
atmospheric and sea ice nullcline is how the conceptual model captures the
effects of the background climate state.

\textbf{Phase B: supercritical case
  (\(\boldsymbol{I_{\mathrm{p}}<I_{\mathrm{B}2}}\)) - Stage 2.} Once the
sea ice has regrown past \(I_{\mathrm{B}2}\), its further regrowth
accelerates substantially, marking the beginning of the second stage in the
supercritical system-wide relaxation process. This is due to the strongly
changing albedo effect around intermediate sea ice cover. The regrowing sea
ice increasingly prevents oceanic heat loss to the atmosphere, which
entails polar atmospheric cooling and initiates warming of the Nordic Seas.
The reduction of the oceanic temperature gradient is closely followed by a
corresponding reduction in the AMOC strength. Since the atmosphere
equilibrates quasi-adiabatically to the declining mutual relaxation rate
\(\gamma(I)\), it cools at the same accelerated rate as the ice recovers
and both atmosphere and sea ice reach stadial configuration within a few
hundred years after the sea ice passed its threshold \(I_{\mathrm{B}2}\).
The oceanic variables follow with some inertia; they exhibit a sustained
relaxation after the faster atmosphere and sea ice have already clearly
transitioned to their stadial configuration. This second stage of the
supercritical relaxation process corresponds to the abrupt
interstadial--stadial transitions observed in the paleoclimate record.


\subsection{Noise-driven interstadials}
\label{sec:noise_results}
\begin{figure*}[t]
\centering
\includegraphics[width = 0.8\textwidth]{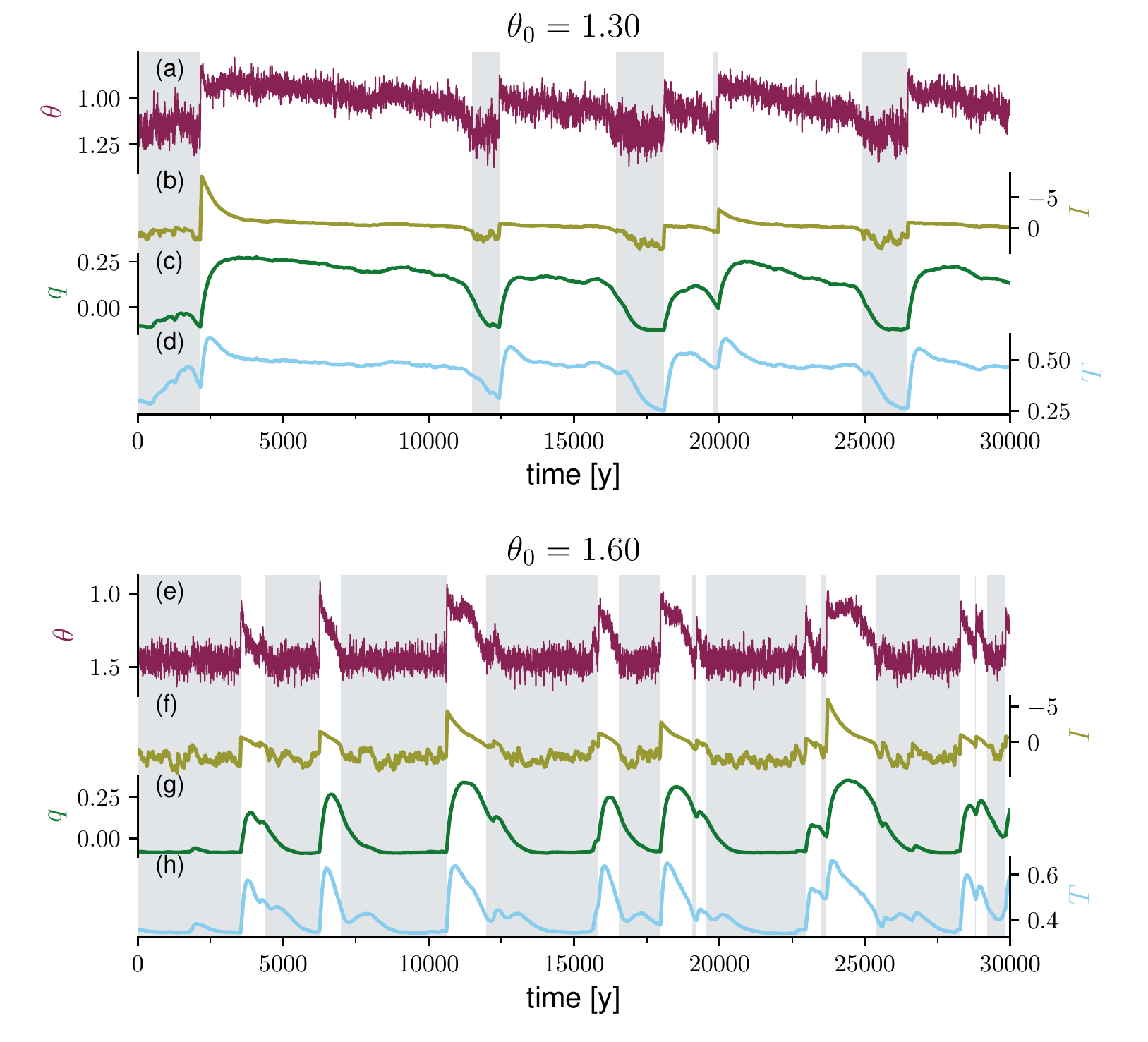}
\caption{Trajectories of the model system defined by
  Eqs.~\ref{eq:1}--\ref{eq:5} driven by the noise scheme as described in
  Sect.~\ref{sec:methods}\ref{sec:noise}, i.e., with non-zero noise
  $\zeta_t$ and $\xi_t$, for \(\theta_{0} = 1.3\) (a--d) and
  \(\theta_{0} = 1.6\) (e--h). The gray shading indicates stadial
  intervals. A DO event is defined by at least 25 consecutive years of sea
  ice cover \(I>I_{\mathrm{B2}}\) within a stadial followed by at least 15
  years of reduced sea ice cover of \(I<I_{\mathrm{B2}}\). Provided that
  the system is in an interstadial state, the reverse interstadial--stadial
  transition occurs when the sea ice regrows past \(I_{\mathrm{c}} = 0.5\)
  and maintains this level in an average over the following 25 years. The
  hysteresis in the definition of climate transitions prevents a jumping
  back and forth between the two states when the sea ice fluctuates close
  to a potential critical threshold and gives rise to well defined climatic
  periods.}
  \label{fig:noise_induced_transition}
\end{figure*}

We now show that the intermittent driving noise $\xi_t$ is able to generate
supercritical sea ice perturbations capable of triggering DO events.
Fig.~\ref{fig:noise_induced_transition} shows trajectories of the dynamics
determined by Eqs.~\ref{eq:1}--\ref{eq:3} and Eqs.~\ref{eq:4} and
\ref{eq:5} under the influence of the driving noises $\xi_t$ and $\zeta_t$
with a constant atmospheric background climatic state $\theta_0$. Overall
there is high visual agreement between simulated \(\theta\) trajectories
(Fig.~\ref{fig:noise_induced_transition}) and the DO cycles recorded in the
NGRIP $\delta^{18}$O data (Fig.~\ref{fig:01}a and c).

For \(\theta_{0}=1.3\) the \(\theta\)-trajectory resembles DO cycles from
the early glacial, with persistent interstadials separated by short
stadials. For \(\theta_{0}=1.6\) the \(\theta\)-trajectory has a greater
similarity with the mid to late glacial, with shorter interstadials and
predominantly stadial conditions. 
Confirming the discussion in the previous section, interstadials last
longer and have a more pronounced plateau in all variables for the smaller
atmospheric background temperature gradient \(\theta_{0} = 1.3\).
Moreover, an increase in the stadial duration can be observed for
larger \(\theta_{0}\).
The colder stadial conditions associated with larger \(\theta_{0}\) imply
an increased distance between the sea ice's stable stadial configuration
\(I_{\mathrm{s}}\) and the critical sea ice threshold \(I_{\mathrm{B}2}\).
Hence, for unchanged driving noise \(\xi_t\), the probability for the noise
to drive the system across \(I_{\mathrm{B}2}\) is reduced for larger
\(\theta_{0}\) and thus the waiting time between two supercritical
stochastic forcing events is higher and the stadials are longer. However,
this effect is attenuated by the fact that DO events can also be triggered
by two pronounced yet subcritical laminar forcing periods in quick
succession. 

The Gaussian atmospheric noise process $\zeta_t$ driving the \(\theta\)
variable blurs the exact timing of DO cooling transitions in agreement with
$\delta^{18}$O ice core records. In the sea ice, however, these transitions
are still fairly distinct. Compared to the deterministic setup, sea ice
perturbations of the same strength yield shorter interstadials. Indeed,
small sea ice fluctuations can disrupt the delicate balance of influences
that yields the slow sea ice regrowth of the meta-stable interstadial and
may easily push the system across the critical threshold of
\(I_{\mathrm{B2}}\). This initiates the final stage of accelerated sea ice
regrowth which ends the interstadial.

Remarkably, our model reproduces several observed irregular features of the
$\delta^{18}$O record, with respect to variability in terms of shape,
duration, and amplitude of DO cycles. The different strengths of the sea
ice perturbations translate into different lengths of the interstadial
intervals for given background conditions \(\theta_{0}\). 
In agreement with the NGRIP \(\delta^{18}\)O record, the shorter
interstadials in the simulation with a colder background climate
(\(\theta_{0} = 1.6\)) lack a clear two-stage cooling. Instead, \(\theta\)
transitions back to the stadial state more continuously compared to the
longer interstadials simulated with warmer a background climate of
\(\theta_{0} = 1.3\). Both trajectories exhibit pronounced perturbations
within stadials towards a warmer Arctic atmosphere that do not develop into
a full interstadial. Perturbations of this kind can also be found in the
NGRIP record (compare for example interstadial 5.1, 16.2, and 21.2 in \cite{Rasmussen2014a}). 


\subsection{Realistic climate background}
\label{sec:transient}

\begin{figure}
  \centering
  \includegraphics[width = 0.5\textwidth]{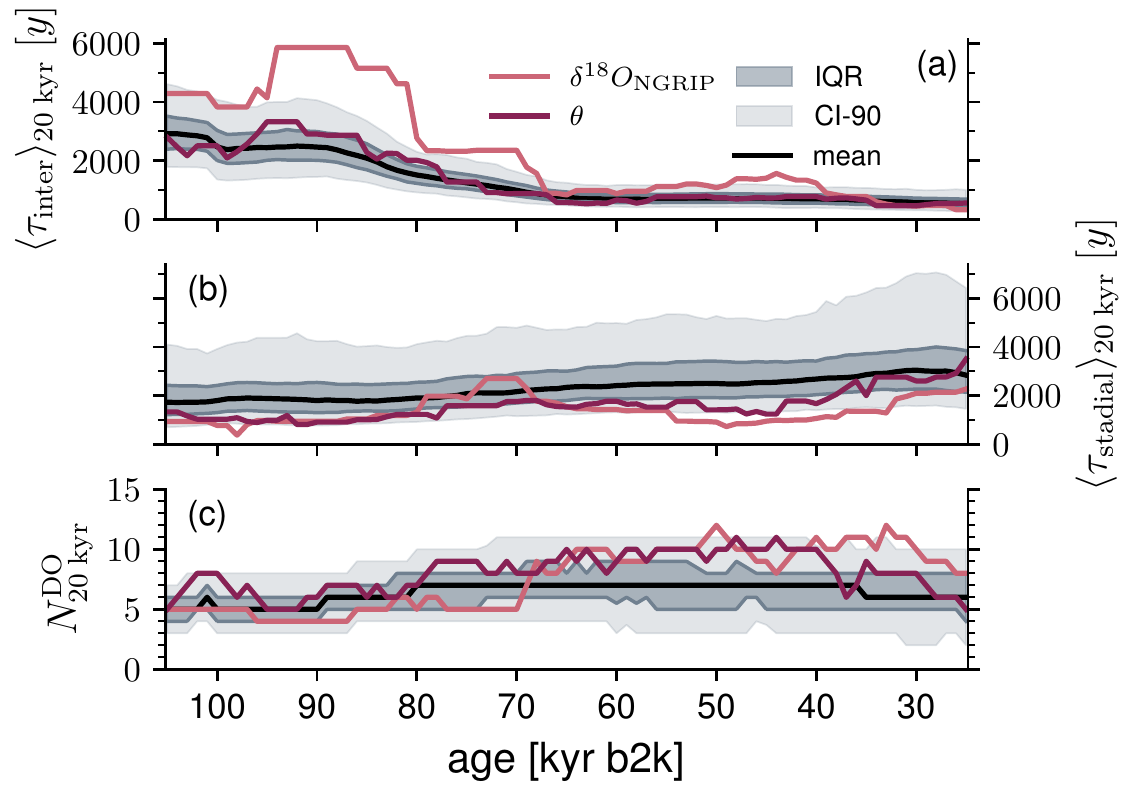}
  \caption{The effect of the changing \(\theta_{0}\) on the interstadial
    and stadial durations. (a) The average duration of all interstadials
    inside running windows of \(20\;\)kyr centered on the respective point
    in time (\(\langle \tau_{\mathrm{inter}}\rangle_{20\;\mathrm{kyr}}\))
    for the proxy data (rose) and the simulation (wine) together with
    corresponding mean (black line), interquartile range (IQR, dark gray)
    and 90\% credibility interval (CI-90, light gray) computed from 1000
    model runs. The \(20\;\)~kyr mean duration takes into account all
    interstadials that are either fully included in the window or that end
    or start within the window. Interstadials that stretch across the
    window boundaries are considered with their full duration. (b) Same as
    (a) for the \(20\;\)~kyr running mean duration of stadials
    (\(\langle\tau_{\mathrm{inter}}\rangle_{20\;\mathrm{kyr}}\)). (c)
    Number of DO events occurring in a 20~kyr running window
    \(N^{\mathrm{DO}}_{20\;\mathrm{kyr}}\) with the same color coding as
    before.}
  \label{fig:08}
\end{figure}
Finally, we run a simulation of the last glacial with the full model
defined by Eqs.~\ref{eq:1}--\ref{eq:5} with a realistically changing
atmospheric background climate according to Eq.~\ref{eq:6} (cf.
Fig.\ref{fig:01}b). Including the temporal variations of the climate
background conditions aligns the simulated stadial and interstadial
durations with those observed in the NGRIP records. The resulting
\(\theta\) trajectory show excellent agreement with the NGRIP
\(\delta^{18}\)O record.

With the linear coupling to the LR04 $\delta^{18}$O stack introduced in
Eq.~\ref{eq:6} the atmospheric background state \(\theta_{0}\) assumes low
values around 1.3 during the early parts of the last glacial and increases
to high values $\theta_0>1.9$ around the last glacial maximum. As shown in
Sect.~\ref{sec:results}\ref{sec:deterministic_dynamics} and
\ref{sec:results}\ref{sec:noise_results}, this leads to longer lasting
interstadials during the early glacial and shorter ones during the late
glacial, with the opposite effect although much less pronounced for
stadials. Hence, the predominance of long lasting interstadials with only
short stadial inceptions in the early glacial is reversed towards the late
glacial (cf. Fig.~\ref{fig:08}). In general, the time scales of the
stadials and interstadials match those observed in the proxy data
throughout the entire last glacial. During the very cold conditions toward
the end of the last glacial, DO events are unlikely but not impossible to
occur in our simulations.


\section{Discussion}

\label{sec:discussion}
Overall, the modelled atmospheric temperature \(\theta\) qualitatively
resembles the NGRIP \(\delta^{18}\)O record over the entire last glacial in
terms of shape and periodicity of DO cycles (compare Fig.~\ref{fig:01}a
with b). The general features of strong (reduced) sea ice cover, weak
(strong) AMOC and warm (cold) Nordic Seas during stadials (interstadials)
are likewise consistently reproduced by the model (compare
Fig.~\ref{fig:01}c--f with g--j).

In principle, the dimensionless units of our model could be translated into
physical units by multiplication with the corresponding scaling factors
(see Appendix). However, since proxies themselves mainly provide
qualitative information on the past we focus on the qualitative analysis of
our model. Above all, we aim to demonstrate a plausible mechanism for DO
cycles and place less emphasis on quantitative accuracy. Nonetheless, it
should be mentioned that multiplying the simulated atmospheric warming of
\(\Delta \theta \sim 0.4\) with a characteristic atmospheric gradient of
\(\theta_{*} \sim 25^{\circ}C\) used for the non-dimensionalization yields
realistic DO warming amplitudes of \(10^{\circ}C\).

\label{sec:model_data_comparison}
We now discuss the similarity between proxy records and our simulations
with focus on the five key characteristics of DO variability as listed in
the introduction, which rely on a multitude of marine, cryosphere and
terrestrial proxy records \cite{Rasmussen1996,Rasmussen2004,Dokken2013,
  Ezat2014,Hoff2016,Lynch-Stieglitz2017,Menviel2020,Sadatzki2020}.

\begin{enumerate}
\item \textbf{Shape of DO cycles:} The general shape of the DO cycles is
  well reproduced by the atmospheric temperature \(\theta\)
  (Fig.~\ref{fig:01}a,c vs. Fig.~\ref{fig:01}b,g). Our model further
  resolves several finer-scale features of the proxy record of the last
  glacial, such as precursor events -- short-lived strong positive
  excursions in the \(\delta^{18}\)O data that do not evolve into full
  interstadials such as the interstadial 5.1 at \(\sim\)31~kyr~b2k or the
  interstadial 16.2 at \(\sim\)58~kyr~b2k or very short stadial inceptions
  similar to the stadial 23.1 at \(\sim\)90~kyr~b2k and the stadial 21.2 at
  \(\sim\)85~kyr~b2k which yield apparent back-to-back interstadials
  \cite{Rasmussen2014a}. Furthermore,
  consistent with the proxy record, several interstadials exhibit a
  continuous trend of moderate cooling back to full stadial conditions
  instead of the more typical abrupt final cooling (compare for example
  interstadials 5.2, 8, and 10 in \cite{Rasmussen2014a}). The slow
  variations in the overall stadial and interstadial levels are reproduced
  by our simulations, however with an apparent mismatch prior to the last
  glacial maximum, where the NGRIP $\delta^{18}$O record shows a persistent
  warming trend (within a stadial climate), while our simulation evolves to
  colder temperatures (cf. Fig.~\ref{fig:08}). This discrepancy is caused
  by the fact that, in contrast to the local warming trend observed in the
  NGRIP record, the LR04 \(\delta^{18}\)O, which drives the background
  climate of our model, indicates a global cooling trend prior to the last
  glacial maximum \cite{Clark2009}. Potentially, this discrepancy could be
  circumvented by coupling \(\theta_{0}\) to orbital parameters instead of
  the background climate.

\item \textbf{Duration of stadials and interstadials:} The modelled
  variability of stadial and interstadial durations is in good agreement
  with the data (cf. Fig.~\ref{fig:08}). In our simulations the coupling of
  the background climate \(\theta_{0}\) to the LR04 benthic $\delta^{18}$O
  introduced in Eq.~\ref{eq:6} causes a gradual transition from a
  predominantly interstadial to a predominantly stadial climate across the
  last glacial due to the mechanisms explained in
  Sects.~\ref{sec:results}\ref{sec:noise_results} and
  \ref{sec:results}\ref{sec:transient}. In the early glacial, our model
  does not generate sufficiently long interstadials. This could be
  circumvented by adjusting the background climate such that the
  interstadial state is weakly stable during this time interval. However,
  this would come at the cost that simulated interstadials would no longer
  show a cooling trend. Between 50 and \(40\;\)kyr b2k our model generates
  slightly too few DO events resulting in a too high stadial duration.

\item \textbf{In-phase sea ice dynamics:} The extensive (reduced) sea ice
  cover during stadials (interstadials) suggested by proxy records
  \cite{Dokken2013, Hoff2016, Sadatzki2019, Sadatzki2020} is well
  reproduced by our model; the sea ice $I$ consistently regrows at a
  moderate rate over the course of the interstadial, before it returns to
  its stadial extent in a final phase of accelerated regrowth marking the
  interstadial--stadial transition. Overall, this behavior agrees with the
  sea ice dynamics in the Nordic Seas across DO cycles as inferred by
  \cite{Sadatzki2020} and others \cite{Dokken2013, Ezat2014, Hoff2016}. While proxy records suggest a continued
  sea ice growth during the early stadial, the modeled sea ice $I$ reaches
  its stadial extent already during the interstadial--stadial transition
  and the regrowth thus does not stretch significantly into the stadial
  phase.
  

\item \textbf{Nordic Seas' temperature inversion:} 
  In our interpretation of the modeled oceanic temperature gradient \(T\)
  we ignore surface waters and regard \(T\) as an indicator of subsurface
  and deep water temperatures. Indeed, our model simulates warm Nordic Seas
  at full stadial conditions (Fig.~\ref{fig:01}j), in line with the proxy
  records \cite{Rasmussen2004, Dokken2013,Ezat2014}. At the beginning of
  interstadials, pronounced cooling sets in, which can be interpreted as a
  continuous convection-driven release of the heat previously stored
  underneath the sea ice cover. Approximately one third into an
  interstadial the subsurface cooling is reversed into a more gentle yet
  persistent warming trend caused by regrowing sea ice and decreasing heat
  loss which accelerates across the transition back to stadial conditions,
  but is sustained in the subsequent stadial. In general, this pattern
  agrees very well with existing paleoclimate proxies \cite{Rasmussen2004, Dokken2013,Ezat2014,Sadatzki2020}. The
  moderate warming that persists over the final two thirds of the
  interstadial can be observed in a very similar manner in the benthic
  $\delta^{18}$O from marine sediment cores indicative of deep ocean
  temperatures (compare Fig.~\ref{fig:01}f with Fig.~\ref{fig:01}j). The
  deep ocean warming is attributed to a gradual reduction of deep
  convection over the course of the interstadial driven by regrowing sea
  ice \cite{Sadatzki2020, Ezat2014}.
  The observed sustained warming of the intermediate and deep ocean into
  stadials is reproduced by our model, where a salinity driven stadial AMOC
  (\(q<0\)) counteracts an existing meridional temperature gradient and the
  sea ice cover prevents heat loss to the atmosphere.
  
\item \textbf{AMOC switches:} The correspondence between strong (weak)
  overturning and interstadial (stadial) climate conditions is widely
  accepted \cite{Broecker1985, Ganopolski2001,Lynch-Stieglitz2017,
    Henry2016, Gottschalk2015, Vettoretti2018, Menviel2020}. However,
  limited quality of the proxy data prevents a more detailed assessment of
  the changes in the AMOC during the course of a typical DO cycle
  \cite{Henry2016, Lynch-Stieglitz2017}. It seems that the AMOC almost
  stopped during Heinrich stadials, while during non-Heinrich stadials it
  probably operated in a weak and shallow mode \cite{Lynch-Stieglitz2017}.
  An AMOC reinvigoration is reported to have happened in synchrony with
  abrupt Greenland warmings, within the limits of dating uncertainties
  \cite{Lynch-Stieglitz2017}. In general terms, simulated changes of the
  AMOC strength agree with this pattern apart from the specific expression
  of Heinrich events, which are not explicitly targeted by our modeling
  setup (c.f. Fig.~\ref{fig:01}e and i).
  From the interstadial onset onward the AMOC strength increases until it
  plateaus somewhere half way through the interstadial. Thereafter a
  weakening trend sets in the strength of which is related to the climate
  background condition and which stretches well into the stadial before the
  AMOC re-assumes its stadial state.

  Notice that in the stable stadial state of the model, the AMOC is in fact
  negative, i.e. its flow is reversed with respect to the modern AMOC.
  Certainly, a complete reversal of the real AMOC seems unphysical. This
  could be circumvented by heuristically adding a constant positive offset
  to the AMOC strength representing, for example, a wind-driven component
  \cite{Vettoretti2022}. However, summarizing earlier findings
  \cite{Rasmussen2004} pointed out that the flow across the
  Iceland-Scotland ridge may in fact have been reversed between stadials
  and interstadials. The at first sight unintended feature of the Stommel
  model could, in fact, reflect a real physical mechanism.


\end{enumerate}

In summary, our four-dimensional model defined by
Eqs.~\ref{eq:1}--\ref{eq:5} reproduces central features of DO variability
in terms of Arctic temperatures $\theta$, Nordic Seas' sea ice cover $I$,
intermediate-to-deep water temperatures $T$, and the meridional overturning
strength $q$. In particular, our modelling results align well with
the characteristics of DO variability inferred from the study of marine
sediment cores from the northern North Atlantic and the Nordic Seas. First
pointed out by \cite{Rasmussen1996, Rasmussen1996a} a sustained inflow of
North Atlantic warm water into the Nordic Seas during stadial periods was
later confirmed and integrated in a conceptual explanation for the
emergence of DO variability by \cite{Rasmussen2004}: The sustained inflow
of warm and salty waters during stadials is subducted under a strong
halocline which eventually separates a cold and fresh surface layer from
the intermediate and deep waters. These experience a gradual warming up to
the point where the growing vertical temperature gradient destabilizes the
stratification and (re)initiates deep convection. This in turn
reinvigorates the AMOC and cools the intermediate to deep waters in the
Nordic Seas. The transition back to the stadials is more or less explicitly
attributed to the prevailing glacial climate background conditions
\cite{Rasmussen2004}. \cite{Dokken2013} and later \cite{Sadatzki2020}
supplemented this framework by providing observational evidence for
extended (reduced) Nordic Seas' ice cover during stadials (interstadials),
highlighting the role of the sea ice as an insulator between atmosphere and
ocean. Our model shows excellent agreement with corresponding proxy
evidence (compare Fig.~\ref{fig:01}) and integrates well into the above
narrative. The insulating effect of the sea ice, which has previously also
been considered by \cite{Boers2018} as a crucial component, is modeled
explicitly in our study and is confirmed to be key to the changes between
stadial and interstadial climates. It is worth mentioning that previous
studies based on conceptual and intermediate complexity models leveraged
changes (or perturbations) in the freshwater forcing as DO triggers instead
of sudden shifts in the atmosphere--ocean heat flux
\cite{Ganopolski2001, Ganopolski2002, Timmermann2003, Menviel2014,
  Roberts2017}. The observed sustained stadial inflow of warm North
Atlantic waters into the intermediate depth Nordic Seas corresponds to a
weak, salinity-driven AMOC in our model.


Importantly, in our model interstadials do correspond to meta-stable states
that inevitably decay back to stadial climate conditions. This excitation
mechanism previously investigated by \cite{Ganopolski2002} and
\cite{Vettoretti2022} adds an alternative view on DO cycles complementing
the more common concepts based on either bi-stability or limit-cycle
behavior \cite{Ditlevsen1999, Timmermann2000, Ditlevsen2005, Livina2010,
  Rial2011, Kwasniok2013, Mitsui2017, Roberts2017, Lohmann2019a}. Our model
does not resolve the exact triggering mechanism and instead relies on
intermittent noise, which may trigger DO events by removing sufficiently
large amounts of sea ice reactivating atmosphere--ocean interaction in the
high latitudes. So far, as a source of this noise, we have proposed either
convective events in the ocean inspired by \cite{Sadatzki2020},
\cite{Dokken2013}, \cite{Rasmussen2004}, and \cite{Vettoretti2018} or
atmospheric anomalies as described by \cite{Kleppin2015} and
\cite{Li2019}. The interplay of both mechanisms, as suggested by
\cite{Sadatzki2020}, might even be better suited to justify the choice of
our driving noise.

Apart from its central role as the DO event trigger, we find that the
driving intermittent process can explain specific details of the NGRIP
$\delta^{18}$O record. The variable strength of the supercritical
perturbations generates variability in terms of the shape and the duration
of interstadials similar to the observations. Similarly, the randomly
distributed waiting times between supercritical perturbations control the
variability of stadial durations. The stochastic nature of the DO trigger
naturally allows for the reproduction of back-to-back interstadials which
lack an extended stadial separating them
and very short interstadials. Furthermore, subcritical forcing events
contribute to the observed larger fluctuations of the stadial Arctic
atmospheric temperatures compared to those during interstadials.
Subcritical events could be interpreted as weaker local convective events
that only entail partial and temporary removal of stadial sea ice with
limited released of oceanic heat. In summary, we find a purely
deterministic mechanism of DO events to be more difficult to reconcile with
the variability of interstadial shapes in the proxy records.

It should be noted that the presence of the three different time scales
\(\tau_{\mathrm{ocean}}, \tau_{\mathrm{ice}}\) and \(\tau_{\mathrm{atm}}\)
is crucial to obtain agreement between the modeled and the observed shape
and duration of DO cycles. In particular, the relatively slow time scale of
the sea ice is required to obtain sufficiently long interstadials. The
other key ingredient for the successful simulation of sustained
interstadial intervals is the proximity of the \(\theta\) and ice
nullclines, which gives rise to an additional dynamics-induced slow time
scale and a meta-stable state. The existence of such a meta-stable state,
whose expression is highly sensitive to background conditions, seems
physically plausible in view of a delicate interplay between northward
oceanic heat transport, the high latitude atmospheric temperatures and the
ice-albedo feedback. The influence of warm Atlantic surface inflow in the
Nordic Seas on the formation of sea ice during interstadials has so far
been neglected in our model, but should be considered in further research.

We acknowledge that the Stommel model represents the AMOC in a drastically
simplified manner which misses import aspects like the AMOC's previously
mentioned dependence on wind forcing \cite[e.g.]{Yang2016, Weijer2019} or
the relation between the AMOC strength and the pycnocline depth
\cite{Gnanadesikan1999, DeBoer2010, Nikurashin2011}. It also neglects the coupling of the AMOC to other ocean basins and in particular the Southern ocean, whose role in the DO variability was emphasized by \cite{Hines2019} and \cite{Thompson2019}. 

Recent simulations with comprehensive climate models were able to reproduce DO-like climate oscillations \cite{Malmierca-vallet2023}. These simulations do consistently reproduce the in-phase sea ice dynamics, the Nordic Seas’ temperature inversion and the AMOC switches as evident from the paleoclimate record and to a higher or lesser degree the general sawtooth-shape of the Greenland temperatures over the course of one cycle. However, these simulations are run under constant background climate conditions and generally produce fairly stable DO cycle periods. They do not recover the large degree of variability of cycle durations and shapes observed in the NGRIP \(\delta^{18}\)O record. 

Targeted experiments have shown that orbital parameters, CO\(_{2}\)
concentration and ice sheet heights affect the cycle period
\cite{Zhang2021, Kuniyoshi2022, Vettoretti2022}. However, it seems unclear
if these models could reproduce the full spectrum of DO cycle variability
in terms of amplitude, period and shape in a transient simulation with
changing ice sheets, orbital parameters and CO\(_{2}\). Here, our study may
serve as a motivation to investigate why comprehensive climate models
produce so regular DO cycles. According to our model, the key ingredient
for generating irregular DO cycles is the stochastic sea ice dynamics which
is driven by an intermittent process. It might be worthwhile exploring if
the models in question underestimate sea ice variability and if this leads
to too stable DO cycles.



\section{Conclusions}
\label{sec:conclusions}
In summary, we have modeled DO cycles across the last glacial period as
state space excursions of an excitable monostable system, resolving the
ocean, sea ice and atmosphere on increasingly fast time scales. DO events
are triggered by a stochastic intermittent process that acts on the stadial
sea ice cover. The associated effect on the atmospheric Arctic temperatures
$\theta$ is consistent with the observed \(\alpha\)-stable noise signature
in the GRIP calcium record \cite{Ditlevsen1999}.
We propose as potential sources for the driving intermittent noise local
and temporary convective instabilities in the stadial stratification of the
Nordic Seas \cite{Singh2014,Jensen2016}, persistent atmospheric anomalies
\cite{Kleppin2015,Li2019}, or combinations thereof \cite{Sadatzki2020}.
Our model reproduces several central aspects of the observed DO cycle
patterns of four climate variables central to the physics of DO cycles: the
typical saw-tooth shape of Arctic atmospheric temperatures
\cite{Johnsen2001, NGRIP2004}, the reduced (extended) sea ice cover during
interstadials (stadials) \cite{Sadatzki2020, Sadatzki2019, Hoff2016,
  Dokken2013}, the strong interstadial AMOC, with sustained northward heat
transport during stadials at a weaker level \cite{Rasmussen1996,
  Rasmussen2004, Dokken2013, Lynch-Stieglitz2017}, and the corresponding
stadial warming of the Nordic Seas \cite{Sadatzki2020, Dokken2013,
  Ezat2014, Rasmussen2004}. Furthermore, detailed aspects of the NGRIP
$\delta^{18}$O record, such as the variability of the interstadial shape
and duration, higher-amplitude stadial fluctuations, and very short
interstadials (stadials) during the colder (warmer) parts of the late
glacial, are reproduced.

We summarize the most important features of our model that provide
plausible mechanistic explanations to the last glacial's millennial-scale
climate variability:

\begin{enumerate}
\item The intensity of ocean--atmosphere heat flux in the high northern
  latitudes is decisive for the state of the North Atlantic climate system.
  The heat flux is controlled by the sea ice. 
  
\item Stochastic removal of the sea ice may abruptly expose the atmosphere
  to the influence of a large oceanic heat reservoir, causing abrupt
  atmospheric warming and oceanic cooling facilitated by reactivated deep
  convection. The latter happens at a much slower rate due to the ocean's
  much larger heat capacity.
  
\item If sea ice is absent the oceanic heat loss to the atmosphere yields a
  strong meridional oceanic temperature gradient which in turn
  drives the AMOC's strong circulation mode.
  
\item In the interstadial configuration the North Atlantic climate system
  is only approximately close to an equilibrium state. Generally prevailing
  cold atmospheric temperatures in the high latitudes entail sea ice
  regrowth, which inevitably drive the system back to the stadial state.

\item The persistence of the interstadial state is highly sensitive to the
  background conditions. 

\item The time-scale separation between ocean, sea ice, and atmosphere
  (from slow to fast) is central to the characteristic shape of Greenland
  interstadials in terms of Greenland temperature changes.

\item DO events are stochastically triggered by a random asymmetric
  intermittent process, mimicking abrupt large sea ice removal events.
  
\end{enumerate}

We hope that our findings provide a helpful conceptual framework for
further investigation of DO variability. We suggest testing the formulated
excitability mechanism in comprehensive climate models by forced removal of
sea ice. In that context, it could also be tested if stronger sea ice
perturbations give rise to longer interstadials in comprehensive models as
suggested by our model. Finally, it seems worthwhile to investigate the sea
ice fluctuations in those models that successfully reproduce DO cycles and
check if their statistics exhibit fat tails in the duration of reduced sea ice extent.

\section*{Acknowledgments}
This is TiPES contribution \#215; the TiPES (`Tipping Points in the
  Earth System') project has received funding from the European Union's
  Horizon 2020 research and innovation programme under grant agreement No.
  820970. NB acknowledges further funding by the Volkswagen Foundation, the
  European Union’s Horizon 2020 research and innovation programme under the
  Marie Sklodowska-Curie grant agreement No. 956170, as well as from the
  German Federal Ministry of Education and Research under grant No.
  01LS2001A.
\bibliographystyle{unsrt}  
\bibliography{bib}

\end{document}


\section*{Supporting information for}
\subsection*{Glacial abrupt climate change as a multi-scale phenomenon resulting from monostable excitable dynamics}
Keno Riechers, Georg Gottwald \& Niklas Boers (2023) \\
\textit{Corresponding Author:} Keno Riechers, riechers@pik-potsdam.decrease
\thispagestyle{fancy}

\paragraph{Detailed derivation of the deterministic part of the model}

The starting point of our model derivation is the classical Stommel model \cite{Stommel1961, Cessi1994}. This models treats the North Atlantic ocean as two separate well mixed boxes, each being characterized by a temperature \(T_{\mathrm{e,p}}\) and a salinity \(S_{\mathrm{e,p}}\). Here the indices e and p refer to the \textit{equatorial} and \textit{polar} box, respectively. In the classical Stommel model, the temperatures in the oceanic boxes relax towards prescribed temperatures of the atmosphere above. Here, we explicitly take into account a response of the atmospheric temperatures \(\theta_{\mathrm{e,p}}\) to the heat fluxes between atmosphere and ocean and correspondingly treat \(\theta_{\mathrm{e,p}}\) as dynamical variables. Figure~2 (main text) shows a schematic illustration of the model model.  

The atmosphere is exposed to net differential heating \(Q_{\mathrm{e,p}}\)
caused by radiative imbalance between equatorial and polar regions. The
diffusive meridional fluxes
\begin{equation}
  \label{eq:7}
  \chi_{\theta} = \gamma_{\theta} (\theta_{\mathrm{e}} - \theta_{\mathrm{p}}), \quad  \chi_{T} = \gamma_{T} (T_{\mathrm{e}} - T_{\mathrm{p}}) \quad \text{and} \quad \chi_{S} = \gamma_{S} (S_{\mathrm{e}} - S_{\mathrm{p}}) 
\end{equation}
counteract existing gradients in both the ocean and the atmosphere,
respectively. The effective heat diffusivities \(\gamma_{\theta}\) and
\(\gamma_{T}\) are given in Watt per Kelvin, so that \(\chi_{\theta, T}\)
denotes a total heat flux across the boundary between the boxes. Similarly,
\(\gamma_{S}\) is given in kilogram per second per PSU so that again, the
flux \(\chi_{S}\) denotes an effective flux of salt.

The atomsphere--ocean heat flux
\begin{equation}
  \label{eq:8}
  \phi_{\mathrm{e,p}} = \gamma_{\mathrm{e,p}} (\theta_{\mathrm{e,p}} - T_{\mathrm{e,p}}) \quad \mathrm{with} \quad \gamma_{\mathrm{e,p}} \propto A_{\mathrm{e,p}}
\end{equation}
is likewise assumed to be proportional to the corresponding temperature
difference between atmosphere and ocean. The flux parameters
\(\gamma_{\mathrm{e,p}}\) are proportional to the corresponding
atmosphere--ocean interface surfaces \(A_{\mathrm{e}}\) and
\(A_{\mathrm{p}}\). The salinities in the oceanic boxes are forced by an
effective freshwater flux \(\sigma_{\mathrm{e,p}}\) caused by an imbalance
between relatively stronger evaporation over the tropics and elevated
precipitation and river runoff in the polar region.
A meridional density gradient \(\Delta \rho = \rho_{\mathrm{e}} - \rho_{\mathrm{p}}\) between equator and pole drives an additional overturning flow
\begin{equation}
  \label{eq:9}
  \psi = k \Delta \rho
\end{equation}
that represents the AMOC in this simple model. 

These considerations lead to the following set of equations:
\begin{subequations}
  \begin{align}
    \dot{\theta}_{\mathrm{e}} &= \frac{1}{\lambda_{\mathrm{e}}}\left(Q_{\mathrm{e}} - \gamma_{\mathrm{e}} (\theta_{\mathrm{e}} - T_{\mathrm{e}}) -  \gamma_{\theta} (\theta_{\mathrm{e}} - \theta_{\mathrm{p}})\right)\\
    \dot{\theta}_{\mathrm{p}} &= \frac{1}{\lambda_{\mathrm{p}}}\left(Q_{\mathrm{p}} - \gamma_{\mathrm{p}} (\theta_{\mathrm{p}} - T_{\mathrm{p}}) + \gamma_{\theta} (\theta_{\mathrm{e}} - \theta_{\mathrm{p}})\right)\\[1em]
    \dot{T}_{\mathrm{e}} &= \frac{1}{\Lambda_{\mathrm{e}}}\left(\gamma_{\mathrm{e}} [\theta_{\mathrm{e}} - T_{\mathrm{e}}] - \gamma_{T} [T_{\mathrm{e}} - T_{\mathrm{p}}]\right) - \frac{1}{V_{\mathrm{p}}} k|\Delta \rho| (T_{\mathrm{e}}-T_{\mathrm{p}})\\
    \dot{T}_{\mathrm{p}} &= \frac{1}{\Lambda_{\mathrm{p}}}\left(\gamma_{\mathrm{p}} [\theta_{\mathrm{p}} - T_{\mathrm{p}}] + \gamma_{T} [T_{\mathrm{e}} - T_{\mathrm{p}}]\right) + \frac{1}{V_{\mathrm{e}}} k|\Delta \rho| (T_{\mathrm{e}}-T_{\mathrm{p}})\\[1em]
    \dot{S}_{\mathrm{e}} &= \frac{1}{V_{\mathrm{e}}}\left(\sigma_{\mathrm{e}} - \gamma_{S} (S_{\mathrm{e}} - S_{\mathrm{p}}) - k|\Delta \rho| [S_{\mathrm{e}}- S_{\mathrm{p}}]\right)\\
    \dot{S}_{\mathrm{p}} &= \frac{1}{V_{\mathrm{p}}}\left(\sigma_{\mathrm{p}} + \gamma_{S} (S_{\mathrm{e}} - S_{\mathrm{p}}) + k|\Delta \rho| [S_{\mathrm{e}}- S_{\mathrm{p}}]\right),
  \end{align}
\end{subequations}
where the heat capacities of the ocean boxes \(\Lambda_{\mathrm{e,p}} = c_{0}\rho_{0}V_{\mathrm{e,p}}\) are given by the product of the specific heat capacity \(c_{0}\), a reference density \(\rho_{T}\) and the box volume \(V_{\mathrm{e,p}}\) (analogously for the atmospheric heat capacities \(\lambda_{\mathrm{e,p}} = c_{\theta}\rho_{\theta}W_{\mathrm{e,p}}\), with \(W_{\mathrm{e,p}}\) denoting the atmospheric box volume). The overturning flow \(\psi\) enters these equations with an absolute value, because the direction of the flow is irrelevant for the exchange of heat and salt between the oceanic boxes. 

Introducing the meridional gradients \(X = X_{\mathrm{e}} - X_{\mathrm{p}}\), e.g. \(\theta = \theta_{\mathrm{e}} - \theta_{\mathrm{p}}\), the equations of motion can be reformulated as follows: 
\begin{subequations}
  \begin{align}
    \label{eq:34}
     \begin{split}
       \frac{\mathrm{d}\theta}{\mathrm{d}t} &=\frac{\lambda_{\mathrm{p}}Q_{\mathrm{e}}-\lambda_{\mathrm{e}}Q_{\mathrm{p}}}{\lambda_{\mathrm{e}}\lambda_{\mathrm{p}}}  -\frac{\gamma_{\mathrm{e}}}{\lambda_{\mathrm{e}}} (\theta_{\mathrm{e}} - T_{\mathrm{e}}) + \frac{\gamma_{\mathrm{p}}}{\lambda_{\mathrm{p}}}(\theta_{\mathrm{p}}-T_{\mathrm{p}}) - \gamma_{\theta}\frac{\lambda_{\mathrm{p}}+\lambda_{\mathrm{e}}}{\lambda_{\mathrm{e}}\lambda_{\mathrm{p}}}\theta \\
       & = \frac{\lambda_{\mathrm{p}}Q_{\mathrm{e}}-\lambda_{\mathrm{e}}Q_{\mathrm{p}}}{\lambda_{\mathrm{e}}\lambda_{\mathrm{p}}}  -\frac{\gamma_{\mathrm{e}}}{\lambda_{\mathrm{e}}} (\theta_{\mathrm{e}} - T_{\mathrm{e}}) + \frac{\gamma_{\mathrm{p}}}{\lambda_{\mathrm{p}}}(\theta-T - \theta_{\mathrm{e}} + T_{\mathrm{e}}) - \gamma_{\theta}\frac{\lambda_{\mathrm{p}}+\lambda_{\mathrm{e}}}{\lambda_{\mathrm{e}}\lambda_{\mathrm{p}}}\theta \\
       & = \frac{\lambda_{\mathrm{p}}Q_{\mathrm{e}}-\lambda_{\mathrm{e}}Q_{\mathrm{p}}}{\lambda_{\mathrm{e}}\lambda_{\mathrm{p}}}  -\frac{\gamma_{\mathrm{e}}\lambda_{\mathrm{p}} + \gamma_{\mathrm{p}}\lambda_{\mathrm{e}} }{\lambda_{\mathrm{e}}\lambda_{\mathrm{p}}} (\theta_{\mathrm{e}} - T_{\mathrm{e}}) + \frac{\gamma_{\mathrm{p}}}{\lambda_{\mathrm{p}}}(\theta-T) - \gamma_{\theta}\frac{\lambda_{\mathrm{p}}+\lambda_{\mathrm{e}}}{\lambda_{\mathrm{e}}\lambda_{\mathrm{p}}}\theta \\
       & \approx - \frac{\gamma_{\theta}}{\lambda_{\mathrm{p}}} \left(\theta - \theta_{0}\right) -\frac{\gamma_{\mathrm{p}}}{\lambda_{\mathrm{p}}}
       (\theta - T) + \frac{\gamma_{\mathrm{e}}\lambda_{\mathrm{p}} + \gamma_{\mathrm{p}}\lambda_{\mathrm{e}} }{\lambda_{\mathrm{e}}\lambda_{\mathrm{p}}} (\theta_{\mathrm{e}} - T_{\mathrm{e}})
     \end{split}
    \\[1em]
    \begin{split}
      \frac{\mathrm{d}T}{\mathrm{d}t}  &= \frac{\gamma_{\mathrm{e}}}{\Lambda_{\mathrm{e}}}(\theta_{\mathrm{e}} - T_{\mathrm{e}})-\frac{\gamma_{\mathrm{p}}}{\Lambda_{\mathrm{p}}}(\theta_{\mathrm{p}} - T_{\mathrm{p}})- \frac{\Lambda_{\mathrm{e}} + \Lambda_{\mathrm{p}}}{\Lambda_{\mathrm{e}}\Lambda_{\mathrm{p}}}\gamma_{T}T
      - \frac{V_{\mathrm{e}} + V_{\mathrm{p}}}{V_{\mathrm{e}}V_{\mathrm{p}}} k|\Delta\rho| T\\
  &= \frac{\gamma_{\mathrm{e}}}{\Lambda_{\mathrm{e}}}(\theta_{\mathrm{e}} - T_{\mathrm{e}})-\frac{\gamma_{\mathrm{p}}}{\Lambda_{\mathrm{p}}}(\theta - T - \theta_{\mathrm{e}} + T_{\mathrm{e}})- \frac{\Lambda_{\mathrm{e}} + \Lambda_{\mathrm{p}}}{\Lambda_{\mathrm{e}}\Lambda_{\mathrm{p}}}\gamma_{T}T
      - \frac{V_{\mathrm{e}} + V_{\mathrm{p}}}{V_{\mathrm{e}}V_{\mathrm{p}}} k|\Delta\rho| T\\      
 & = \frac{\Lambda_{\mathrm{e}} \gamma_{\mathrm{p}}+ \Lambda_{\mathrm{p}}\gamma_{\mathrm{e}}}{\Lambda_{\mathrm{e}}\Lambda_{\mathrm{p}}}(\theta_{\mathrm{e}}-T_{\mathrm{e}})-\frac{\gamma_{\mathrm{p}}}{\Lambda_{\mathrm{p}}}(\theta - T)- \frac{\Lambda_{\mathrm{e}} + \Lambda_{\mathrm{p}}}{\Lambda_{\mathrm{e}}\Lambda_{\mathrm{p}}}\gamma_{T}T
 - \frac{V_{\mathrm{e}} + V_{\mathrm{p}}}{V_{\mathrm{e}}V_{\mathrm{p}}} k|\Delta\rho| T\\
  & \approx -\frac{\gamma_{\mathrm{p}}}{\Lambda_{\mathrm{p}}}\left(\theta - T\right)- \left(\frac{\gamma_{T}}{\Lambda_{\mathrm{p}}}
    - \frac{k|\Delta\rho|}{V_{\mathrm{p}}} \right) T + \frac{\Lambda_{\mathrm{e}} \gamma_{\mathrm{p}}+ \Lambda_{\mathrm{p}}\gamma_{\mathrm{e}}}{\Lambda_{\mathrm{e}}\Lambda_{\mathrm{p}}}(\theta_{\mathrm{e}}-T_{\mathrm{e}})
\end{split}
\\[1em]
    \begin{split}
    \frac{\mathrm{d}S}{\mathrm{d}t}  &= \frac{V_{\mathrm{p}}\sigma_{\mathrm{e}} + V_{\mathrm{e}}\sigma_{\mathrm{p}}}{V_{\mathrm{e}}V_{\mathrm{p}}} -\frac{V_{\mathrm{e}} + V_{\mathrm{p}}}{V_{\mathrm{e}}V_{\mathrm{p}}} \left( \gamma_{S}- k|\Delta\rho|\right) S\\
  &\approx \frac{V_{\mathrm{p}}\sigma_{\mathrm{e}} + V_{\mathrm{e}}\sigma_{\mathrm{p}}}{V_{\mathrm{e}}V_{\mathrm{p}}} - \frac{1}{V_{\mathrm{p}}}\left( \gamma_{S}- k|\Delta\rho|\right) S, 
      \end{split}
  \end{align}
\end{subequations}
with
\(\theta_{0} =
\frac{\lambda_{\mathrm{p}}}{\gamma_{\theta}}\frac{\lambda_{\mathrm{p}}Q_{\mathrm{e}}-\lambda_{\mathrm{e}}Q_{\mathrm{p}}}{\lambda_{\mathrm{e}}\lambda_{\mathrm{p}}}\).
Here, we have used the fact that \(W_{\mathrm{e}}\gg W_{\mathrm{p}}\) and
\(V_{\mathrm{e}}\gg V_{\mathrm{p}}\) which implies
\(\lambda_{\mathrm{e}}\gg\lambda_{\mathrm{p}}\) and
\(\Lambda_{\mathrm{e}}\gg\Lambda_{\mathrm{p}}\).

Following \cite{Stommel1961} we express the oceanic densities in terms of
a linearized equation of state
\begin{subequations}
  \begin{equation}
    \label{eq:7}
    \rho_{\mathrm{e,p}}  = \rho_{0} (1 - \alpha_{T}T_{\mathrm{e,p}} + \beta_{S}S_{\mathrm{e,p}}),
  \end{equation}
  \begin{equation}
    \Delta \rho = \rho_{0}(\beta_{S}S - \alpha_{T}T)
  \end{equation}
\end{subequations}
with the reference density \(\rho_{0} = 1029\;\mathrm{kg/m}^{3}\) and the thermal expansion and haline contraction coefficients \(\alpha_{T} = 0.17 \cdot 10^{-3}\;\mathrm{K}^{-1}\) and \(\beta_{S} = 0.75 \cdot 10^{-3} \mathrm{psu}^{-1}\), respectively.

We assume that the equatorial atmosphere and ocean are close to a thermal
equilibrium with respect to each other with \(\theta_{\mathrm{e}}\sim T_{\mathrm{e}}\). This allows us to neglect
contributions of the form \([\theta_{\mathrm{e}}- T_{\mathrm{e}}]\). After rescaling all temperatures and the salinity according to
\begin{equation}
  T' = \frac{T}{\theta_{*}},\quad \theta' = \frac{\theta}{\theta_{*}},\quad  \theta'_{0} = \frac{\theta_{0}}{\theta_{*}} =\mathcal{O}(1)\quad \mathrm{and}\quad S' = \beta_{S}S / \alpha_{T}\theta_{*},
\end{equation}
with \(\theta_{*} = 25^{\circ}C\) denoting a typical atmospheric meridional
temperature gradient, the equations of motion simplify to
\begin{subequations}
  \begin{align}
    \frac{\mathrm{d}\theta'}{\mathrm{d}t} & \approx -       \frac{\gamma_{\theta}}{\lambda_{\mathrm{p}}}\left(\theta' - \theta'_{0}\right) -\frac{\gamma_{\mathrm{p}}}{\lambda_{\mathrm{p}}}                            \left(\theta' - T' \right) \\[1em]
    \frac{\mathrm{d}T'}{\mathrm{d}t}  & \approx -\frac{\gamma_{\mathrm{p}}}{\Lambda_{\mathrm{p}}}\left(\theta' - T'\right)- \left(\frac{\gamma_{T}}{\Lambda_{\mathrm{p}}} - \frac{k\rho_{0}\alpha_{T}\theta_{*}}{V_{\mathrm{p}}} |S' - T'| \right) T'\\[1em]
    \frac{\mathrm{d}S'}{\mathrm{d}t}  &\approx \frac{V_{\mathrm{p}}\sigma_{\mathrm{e}} +
  V_{\mathrm{e}}\sigma_{\mathrm{p}}}{V_{\mathrm{e}}V_{\mathrm{p}}}
\frac{\beta_{S}}{\alpha_{T}\theta_{*}} - \left( \frac{\gamma_{S}}{V_{\mathrm{p}}}- \frac{k\rho_{0}\alpha_{T}\theta_{*}}{V_{\mathrm{p}}}|S' - T'| \right) S'.
  \end{align}
\end{subequations}
Since neither \(\theta_{*}\) nor
\(\sigma_{\mathrm{eff}}\) do depend on time in our model and since the
meridional fluxes \(\chi_{\theta}\) and \(\chi_{T}\) impact the polar boxes
much stronger than the equatorial boxes according to their respective
heat capacities, temporal changes in the gradients are governed by changes
in the polar quantities.

According to \cite{Cessi1994}, we interpret the quantities
\(\gamma_{\mathrm{\theta}} / \lambda_{\mathrm{p}} = \tau_{\theta}^{-1}\),
\(\gamma_{\mathrm{p}}/\Lambda_{\mathrm{p}} = \tau_{\mathrm{r}}^{-1}\),
\(\gamma_{T}/\Lambda_{\mathrm{p}} = \tau_{\mathrm{d}}^{-1}\) and
\(k\rho_{0}\alpha_{T}\theta_{*} / V_{\mathrm{p}} = \tau_{\mathrm{a}}^{-1}\)
as the time scales of atmospheric diffusion, the relaxation of the oceanic
temperatures against the atmospheric ones, the oceanic horizontal diffusion
and the oceanic advection, respectively. Some formulations of the Stommel
model rely on two separate time scales for heat and salt diffusion
\cite[e.g.]{Lohmann2019a}. However, following \cite{Stommel1961} and
\cite{Cessi1994} we assume that the time scale of haline diffusion with
\(\gamma_{S}/V_{\mathrm{p}} = \tau_{S}^{-1}\) is the same as that of
oceanic heat diffusion \(\tau_{S} = \tau_{\mathrm{d}}\). Multiplying all
equations with the ocean diffusion time scale \(\tau_{\mathrm{d}}\) and
omitting the primes yields
\begin{subequations}
  \label{eq:xStommel}
  \begin{align}
    \tau_{\mathrm{d}}\frac{\mathrm{d}\theta}{\mathrm{d}t} & \approx -       \frac{\tau_{\mathrm{d}}}{\tau_{\theta}}\left(\theta - \theta_{0}\right) -\frac{\tau_{\mathrm{d}}}{\tau_{\mathrm{r}}}\frac{\Lambda_{\mathrm{p}}}{\lambda_{\mathrm{p}}}                            \left(\theta- T \right) \\[1em]
    \tau_{\mathrm{d}} \frac{\mathrm{d}T}{\mathrm{d}t}  & \approx -\frac{\tau_{\mathrm{d}}}{\tau_{\mathrm{r}}}\left(\theta - T\right)- \left(1 - \frac{\tau_{\mathrm{d}}}{\tau_{\mathrm{a}}} |S - T| \right) T\\[1em]
\tau_{\mathrm{d}} \frac{\mathrm{d}S}{\mathrm{d}t}  &\approx \sigma_{\mathrm{eff}} - \left(1 - \frac{\tau_{\mathrm{d}}}{\tau_{\mathrm{a}}}|S - T| \right) S.
  \end{align}
\end{subequations}
with
\(\sigma_{\mathrm{eff}} = \frac{V_{\mathrm{p}}\sigma_{\mathrm{e}} +
  V_{\mathrm{e}}\sigma_{\mathrm{p}}}{V_{\mathrm{e}}V_{\mathrm{p}}}
\frac{\beta_{S}\tau_{\mathrm{d}}}{\alpha_{T}\theta_{*}}\).

\paragraph{Estimation of the parameters.}
In our choice of parameters we deviate strongly from the values provided by
\cite{Cessi1994}. First, \cite{Cessi1994} assumes the same volume for the
equatorial and polar oceanic boxes. This is given the product of the
typical depth \(H = 4.5\;\mathrm{km}\), zonal extent
\(\delta_{w} = 300\;\mathrm{km}\) and meridional extent
\(L = 8250\;\mathrm{km}\). Here, we adopt the volume specifications used by
\cite{Park1999}:
\(V = V_{\mathrm{e}} + V_{\mathrm{p}} =
5000\;\mathrm{km}\cdot5000\;\mathrm{km}\cdot4\;\mathrm{km}\cdot\) and
\(V_{\mathrm{e}} \simeq 9 V_{\mathrm{p}} \) which is also in line with a
Nordic Seas surface of \(2.5 \cdot 10^{6}\;\mathrm{km}^{2}\) indicated in
\cite{Drange2005}. Regarding the time scale \(\tau_{\mathrm{r}}\) we argue
that the relaxation involves vertical heat transport within the oceanic
boxes. Clearly, in a \(4\;\mathrm{km}\) deep ocean basin the temperature is
a function of the depth. However, we cannot resolve the resulting
temperature profile in our two-box framework. In particular, our boxes
neglect the dynamics at the ocean's surface layer where complex processes
like freezing and melting and high susceptibility to atmospheric forcing
play an important role. We consider a depth of \(D = 800\;\mathrm{m}\) to
estimate \(\tau_{\mathrm{r}}\), because our model variable
\(T_{\mathrm{p}}\) shall reflect changes in the intermediate to deep Nordic
Seas temperatures \cite{Sadatzki2020}. Together with a typical vertical
diffusivity of \(\kappa_{z} = 1\cdot10^{-4}\;\mathrm{m}^{2}s^{-1}\)
\cite{Vettoretti2022} we obtain
\begin{equation}
  \label{eq:10}
  \tau_{\mathrm{r}} = \frac{D^{2}}{\kappa_{z}} = \frac{1000^{2}\;\mathrm{m}^{2}}{1\cdot10^{-4}\;\mathrm{m}^{2}\;\mathrm{s}^{-1}} = 203 \;\mathrm{years}.
\end{equation}
To estimate the diffusive time scale we use an ocean horizontal diffusivity of \(\kappa_{h} = 1000\;\mathrm{m}^{2}\mathrm{s}^{-1}\) \cite{Cessi1994} which yields
\begin{equation}
  \label{eq:11}
  \tau_{\mathrm{d}} = \frac{L^{2}}{\kappa_{h}} = 793\;\mathrm{years}, 
\end{equation}
with \(L=5000\;\mathrm{km}\) according to the volume specification. 

Given that our polar oceanic box is meant to represent the Nordic Seas, the
advective time scale is given by the ratio between the box volume and the
AMOC's volume transport into the Nordic Seas. According to
\cite{Vettoretti2015} the AMOC inflow into the Nordic seas was
around 3 Sverdrup
(\(1 \mathrm{Sv} = 1\cdot10^{6}\;\mathrm{m}^{3}\mathrm{s}^{-1}\)) during
stadials and 
\begin{equation}
  \label{eq:12}
  \tau_{\mathrm{a}} = \frac{V/10}{3\;\mathrm{Sv}} = 106\;\mathrm{years}
\end{equation}
serves as a rough estimate for the advective time scale in our model.

Finally, for the atmospheric relaxation time scale we choose a value of
\(\tau_{\theta} = 55\;\mathrm{days}\) which ranges between synoptic and
seasonal time scales. This choice guarantees that in the absence of sea ice
the atmospheric gradient \(\theta\) is determined to equal contributions by
the oceanic temperature gradient \(T\) and the background gradient
\(\theta_{0}\) as we will show later.

To compute the ratio between the heat capacities of the high latitude oceanic and atmospheric boxes, we need to specify the volume of the polar atmosphere \(W_{\mathrm{p}}\). We identify the polar atmospheric box with a volume that extends from 60\(^{\circ}\)N to 80\(^{\circ}\)N and from  45\(^{\circ}\)W to 25\(^{\circ}\)E. The height of the troposphere over this area is approximately \(6\;\mathrm{km}\) such that
\begin{equation}
  \begin{split}
    W_{\mathrm{p}} &= \int_{-45^{\circ}}^{25^{\circ}} d\phi  \int_{10^{\circ}}^{25^{\circ}} \sin(\theta) d\theta \int_{R_{0}}^{R_{0} + 6\;\mathrm{km}} r^2 dr \\
   & =-  \phi|_{-45^{\circ}}^{25^{\circ}} \cdot \cos(\theta)|_{10^{\circ}}^{25^{\circ}} \cdot 1/3 r^{3}|_{R_{0}}^{R_{0} + 6\;\mathrm{km}}\\
&\approx 23.3\cdot 10^{6}\;\mathrm{km}^{3}, 
  \end{split}
\end{equation}
where we have used the polar radius of the Earth \(R_{0} = 6357\;\mathrm{km}\). 
The total heat capacity follows by multiplying the volume with a typical density of air  \(\rho_{\theta} = 1.2985 \cdot 1000^{3}\;\mathrm{kg}\;\mathrm{km}^{-3}\) and its specific heat capacity \(c_{\theta} =  1035\;\mathrm{J}\;\mathrm{kg}^{-1}\;\mathrm{K}^{-1}\) :
\begin{equation}
  \lambda_{\mathrm{p}} = \rho_{\theta} c_{\theta} W_{\mathrm{p}}  \approx 3\cdot10^{19}\;\mathrm{J}\;\mathrm{K}^{-1}.
\end{equation}

For the ocean, the above specifications imply a volume of the polar box of
\begin{equation}
  V_{\mathrm{p}} = V / 10 = 5000\;\mathrm{km} \cdot5000\;\mathrm{km} \cdot 4\;\mathrm{km} = 10\cdot 10^{6}\;\mathrm{km}^{3}
\end{equation}  
yielding an oceanic heat capacity of 
\begin{equation}
  \Lambda_{p} = \rho_{0} c_{0} V_{\mathrm{p}}  \approx 4 \cdot 10^{22}\;\mathrm{J}\;\mathrm{K}^{-1}, 
\end{equation}
with the oceanic reference density of \(\rho_{0} = 1029 \cdot 1000^{3}\;\mathrm{kg}\;\mathrm{km}^{-3}\) and its specific heat capacity \(c_{0} = 3.9 \cdot 10^{3}\;\mathrm{J}\;\mathrm{kg}^{-1}\;\mathrm{K}^{-1}\). The ratio of atmospheric and oceanic heat capacities is thus given by \(\Lambda_{\mathrm{p}}  / \lambda_{\mathrm{p}} = 1323 \). Dividing the atmospheric equation of motion by this ratio and introducing the time scale \(\tau_{\mathrm{d}}\lambda_{\mathrm{p}} / \Lambda_{\mathrm{p}} = 0.6\;\mathrm{years}\) immediately reveals a clear time scale separation between atmosphere and ocean.

Finally, we need to specify the effective salinity forcing \(\sigma_{\mathrm{eff}}\). Typically, this value is considered a
bifurcation parameter in applications of the Stommel model
\cite{Stommel1961, Park1999, Roberts2017, Lohmann2021a}. Here we keep the value of \(\sigma_{\mathrm{eff}}\) fixed at 0.7, which is on the lower end of the commonly considered parameter range.

In place of \(\sigma_{\mathrm{eff}}\) we vary \(\tau_{\mathrm{r}}\), which
will give rise to a similar bifurcation structure. We argue that the flux
\(\phi_{\mathrm{p}}\) or equivalently the flux parameter
\(\gamma_{\mathrm{p}}\) is a function of the Nordic Seas' sea ice cover
which, if present, acts as an insulator between atmosphere and ocean. The
value \(\tau_{\mathrm{r}} = 202\;\mathrm{years}\) specified above shall
correspond to ice free conditions. With sea ice being the fourth dynamical
variable of our model, the rate \(\gamma_{\mathrm{p}}(I)\) and
correspondingly \(\tau_{\mathrm{r}}\) is subject to dynamic changes. This
mechanism will be described in detail further below and for now we
concentrate on the solutions of Eqs.~\eqref{eq:xStommel} in dependence on a
variations of \(\gamma_{\mathrm{p}}\propto \tau_{\mathrm{d}}/\tau_{\mathrm{r}}\). For sake of readability we introduce the parameters \(\tau_{\mathrm{atm}} = \tau_{\mathrm{d}}\lambda_{\mathrm{p}} / \Lambda_{\mathrm{p}} \approx 0.6\), \(\tau_{\mathrm{ocean}} = \tau_{\mathrm{d}} \approx 800\), \(\Gamma = (\tau_{\mathrm{d}}\lambda_{\mathrm{p}}) / (\tau_{\theta} \Lambda_{\mathrm{p}}) \approx 4\), \(\gamma = \tau_{\mathrm{d}} / \tau_{\mathrm{r}} \approx 4\) and \(\eta = \tau_{\mathrm{d}} / \tau_{\mathrm{a}} \approx 7.5\) and rewrite Eqs.~\eqref{eq:xStommel} as
\begin{subequations}
  \label{eq:xStommel_short}
  \begin{align}
    \tau_{\mathrm{atm}}\frac{\mathrm{d}\theta}{\mathrm{d}t} & \approx - \Gamma \left(\theta - \theta_{0}\right) - \gamma \left(\theta- T \right) \\[1em]
    \tau_{\mathrm{ocean}} \frac{\mathrm{d}T}{\mathrm{d}t}  & \approx -\gamma\left(\theta - T\right)- \left(1 - \mu |S - T| \right) T\\[1em]
\tau_{\mathrm{ocean}} \frac{\mathrm{d}S}{\mathrm{d}t}  &\approx \sigma_{\mathrm{eff}} - \left(1 - \mu|S - T| \right) S.
  \end{align}
\end{subequations}

Let \(q = T-S\) denote the non-dimensionalized meridional density gradient. From \(\psi = k \Delta \rho\) and \(k = V_{\mathrm{p}} / (\tau_{\mathrm{a}}\rho_{0}\alpha_{T}\theta_{*})\) it follows that
\begin{equation}
  \label{eq:15}
  \psi = \frac{V_{\mathrm{p}}}{\tau_{\mathrm{a}}\rho_{0}\alpha_{T}\theta_{*}}   \alpha_{T}\theta_{*}\rho_{0} q = \frac{V_{\mathrm{p}}}{\tau_{\mathrm{a}}} q = \psi_{0} q.
\end{equation}
So the non-dimensional \(q\) indicates the strength of the AMOC in units of
\(\psi_{0} = 3\;\mathrm{Sv}\). Since \(|T-S|\) is generally less than 1,
this will effectively yield unrealistically small values for the AMOC.
Hence, the model is not self-consistent in the sense that assuming a
typical value for the AMOC transport to estimate the time scale
\(\tau_{\mathrm{adv}}\) implies that the modelled AMOC strength will always
be less than the considered typical AMOC strength. However, since the
available paleoclimate data only provides qualitative information on past
AMOC changes, we correspondingly consider \(q\) as a qualitative indicator
for the AMOC behaviour attaching limited importance to quantitative
correctness.

\paragraph{Dynamics of the extended Stommel model.}
\begin{figure}
  \centering
  \includegraphics[width = 0.5\textwidth]{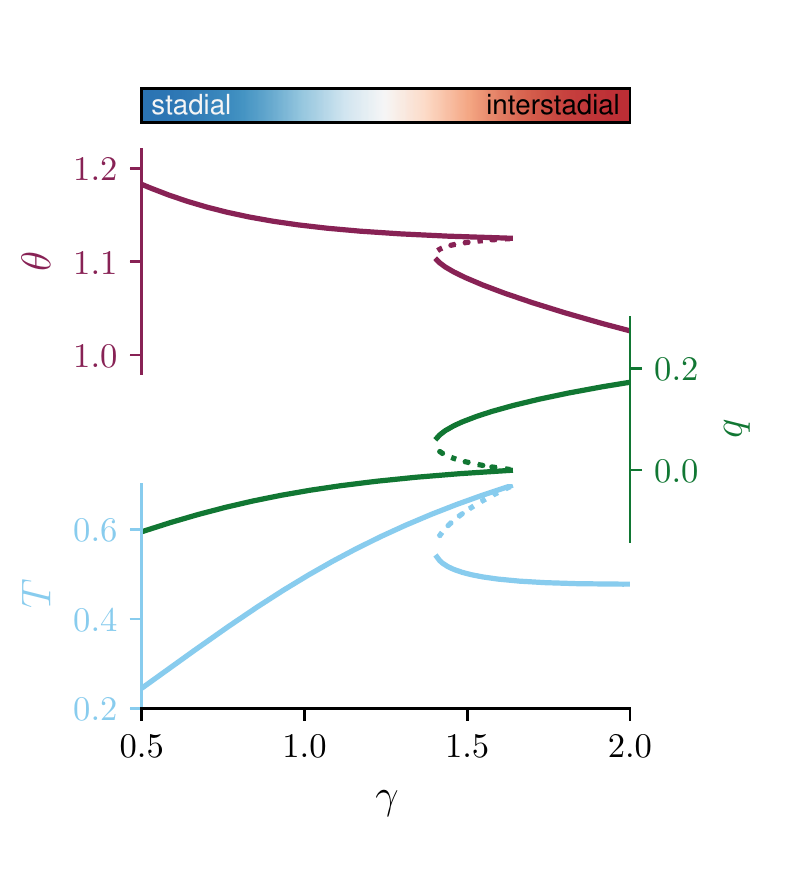}
  \caption{Bifurcation diagram of the extended Stommel model defined by
    Eqs.~\ref{eq:xStommel_short}. The mutual relaxation rate \(\gamma\)
    acts as a control parameter with two bifurcations occurring at
    $\gamma_{c1} \approx 1.4$ and $\gamma_{c2} \approx 1.63$. Solid
    (dashed) lines represent stable (unstable) branches. The atmospheric
    meridional temperature gradient $\theta$ declines with increasing heat
    exchange rate \(\gamma\) due to the action of the ocean on the
    atmosphere. The stable branch of the model that is associated with
    higher \(\gamma\) is commonly referred to as the strong mode
    (temperature driven), while the stable branch associated with lower
    \(\gamma\) is called the weak mode (salinity driven). For low
    \(\gamma <0.8\) all model variables assume a stadial configuration,
    while for \(\gamma >1.7\) they assume an interstadial configuration.
    Here, the atmospheric background climate is set to $\theta_0 = 1.3$ and
    all other parameters are as given in Tab.~1 in the main
    text.}
  \label{fig:gamma_bifurcation}
\end{figure}
We will later introduce explicit dependencies of \(\theta_{0}\) and
\(\gamma\) on the temporarily varying background climate and the sea ice
state, respectively. As a first step, we now discuss the dynamics of the
extended Stommel model defined by Eqs.~\ref{eq:xStommel_short} for a
constant background and a prescribed mutual relaxation rate $\gamma$.
Figure~\ref{fig:gamma_bifurcation} shows the model's bifurcation diagram
with $\gamma$ acting as a bifurcation parameter. As the standard Stommel
model, the extended model has two stable fixed point branches. For
\(1.4<\gamma<1.63\) the system is bistable with an unstable branch
separating the two stable ones. The lower branch is associated with a weak
mode of overturning (smaller \(|q|\)) while the upper branch represents a
strong mode (larger \(|q|\)), with the former being salinity-driven
(\(S>T\)) and the latter being temperature-driven (\(T>S\)). We note that
the increase of \(q\) with increasing \(\gamma\) is stronger in the
standard Stommel model because in the extended model a strengthening of the
AMOC weakens the atmospheric temperature gradient and thus its own driving force.

Given the interpretation of the model variables established in the main
text, with \(\theta\) corresponding to Arctic atmospheric temperatures,
\(T\) representing the Nordic Seas' deep and intermediate water
temperatures and \(q\) indicating the AMOC strength, we can identify
stadial and interstadial climate conditions with different configurations
of the extended Stommel model.
The atmosphere--ocean heat mutual relaxation rate $\gamma$ controls the
state of the high latitude climate (Fig.~\ref{fig:gamma_bifurcation}).
Varying \(\gamma\) from low to high values yields qualitative changes in
all three model variables that consistently match the changes of the true
climate system observed between stadial and interstadial periods. Low
values of \(\gamma \lesssim 0.8\) are associated with a weak AMOC state
(lower \(|q|\) as compare to the strong stable AMOC branch), cold polar
atmosphere (large \(\theta\)) and warm Nordic Seas (small \(T\)), while
high values of \(\gamma \gtrsim 1.7\) reverse this configuration. This
allows us to interpret the stable states of the extended Stommel model at
low and high \(\gamma\) as stadial and interstadial states, respectively.

Although higher values of \(\gamma\) entail enhanced oceanic northward heat
transport, the Nordic Seas are colder in this configuration because of the
stronger release of heat into the atmosphere. In the stadial state (small
values of \(\gamma\)) the model shows sustained northward heat transport
provided by the salinity-driven weak AMOC. However, with the corresponding
smaller atmosphere--ocean heat flux (smaller $\gamma$) even a reduced
northward heat transport warms the northern ocean as it hardly looses any
heat to the atmosphere. This model behavior is in line with the findings by
\cite{Rasmussen2004}, \cite{Dokken2013} and \cite{Ezat2014} who report a
sustained warm water inflow into the Nordic seas at intermediate depth
during stadials. Subsurface warming at high latitudes was also reproduced by complex modelling studies \cite{Vettoretti2018, Kuniyoshi2022}. 


\paragraph{Sea Ice component.}
The sea ice variable \(I\) represents the sea ice cover in the polar box
(i.e. in the Nordic Seas). Acting as an insulator, the sea ice controls the
heat flux \(\phi_{\mathrm{p}}\) between the polar atmosphere and ocean
\cite{Dokken2013,Boers2018}. In the concise model formulation
(Eqs.~\eqref{eq:xStommel_short}), this control may effectively be expressed
by introducing an explicit dependency of the model parameter \(\gamma\) on
the sea ice variable \(I\). To model the sea ice dynamics, we adopt the
seasonally averaged version of the \cite{Eisenman2012} sea ice model
introduced by \cite{Lohmann2021a}:
\begin{equation}
  \label{eq:A16}
  \dot{I} = \Delta \tanh\left(\frac{I}{h}\right)
  - R_{0}H(I)\;I + L - B\;I,
\end{equation}
where $H(I)$ denotes the Heaviside function. The first term represents
the ice-albedo feedback to the incoming solar short-wave radiation. The sea
ice transport, which is absent in open ocean conditions ($I<0$), is
controlled by \(R_{0}\). The term \(L-BI\) describes the change of sea ice
due to the net outgoing longwave radiation (OLR) according to a linearized
Stefan Boltzmann law. The incoming longwave radiation depends on the
atmospheric temperature $\theta$. For the derivation of the non-dimensionalized equation please see \cite{Eisenman2012}. 

In order to incorporate dynamic changes of the polar atmospheric
temperature in Eq.~\ref{eq:A16} and to couple the sea ice model with the
extended Stommel model, we assume a linear relationship between the net
incoming (or outgoing) longwave radiation and the atmospheric temperature
and write
\begin{equation}
 \label{eq:Acoupling_theta_I}
 L = L_{0} + L_{1} (\theta - \theta_{*}), 
\end{equation}
where \(\theta - \theta_{*}\) denotes deviations from the typical
atmospheric temperature gradient. Since larger \(\theta\) corresponds to a
colder polar atmosphere with positive effect on the sea ice growth,
\(L_{1}\) is positive.
Inserting Eq.~\ref{eq:Acoupling_theta_I} into Eq.~\ref{eq:A16} finally
leads to our sea ice model given by Eq.~6 in the main text
\begin{equation}
\label{eq:A4}
\tau_{\mathrm{ice}}\dot{I} = \Delta \tanh\left(\frac{I}{h}\right) - R_{0}H(I)\;I - L_{0} + L_{1}\theta - L_{2}\;I + \xi_t,
\end{equation}
yet without the stochastic forcing $\xi_t$. We ignore influences of the
intermediate to deep ocean temperatures on the sea ice and instead assume that the ocean's surface layer is governed by the atmospheric temperatures.  

Notice that the extended Stommel model is formulated such that energy is
conserved. Incorporating the sea ice model into the Stommel model while
respecting conservation of energy would require to resolve different layers
of the polar ocean box and in particular the fluxes into and out of the
surface layer. Since the heat capacity of the oceans surface layer is
relatively small compared to the heat capacities of the considered oceanic
and atmospheric boxes, we may ignore the exact amount of energy that
enters and leaves this layer.


In the original version of the \cite{Eisenman2012} model the sea ice
variable \(I\) describes the ice thickness over an isolated ocean column of
small spatial extent with horizontally homogeneous temperature. In
particular, negative \(I\) corresponds to ice free conditions
\cite{Eisenman2012}. Given the large spatial extend of the polar box, the
homogeneity assumption does not hold in our application. Instead, we
interpret the variable \(I\) as a stylized representation of the sea ice
over the Nordic seas and its impact on the atmosphere--ocean heat flux. In
this context, we introduce the time scale \(\tau_{\mathrm{ice}}\) to
express that \(I\) reflects slow changes of the annually averaged total sea
ice volume instead of the fairly rapid dynamics of growing or melting sea
ice at an individual point in space. For given climatic conditions sea ice
equilibrates fairly rapidly, meaning that it either forms or melts in a
single season. Multiyear sea ice does usually not exceed an age of roughly
10 years setting an upper bound on typical relaxation time scale of sea ice
with respect to changing climatic conditions at a given location. The much
larger time scale of \(\tau_{\mathrm{ice}} = 200\;\)years chosen for our
study corresponds to the time scale of changes in the multi-year sea
ice edge.



\begin{figure}
  \centering
  \includegraphics[width = 0.5\textwidth]{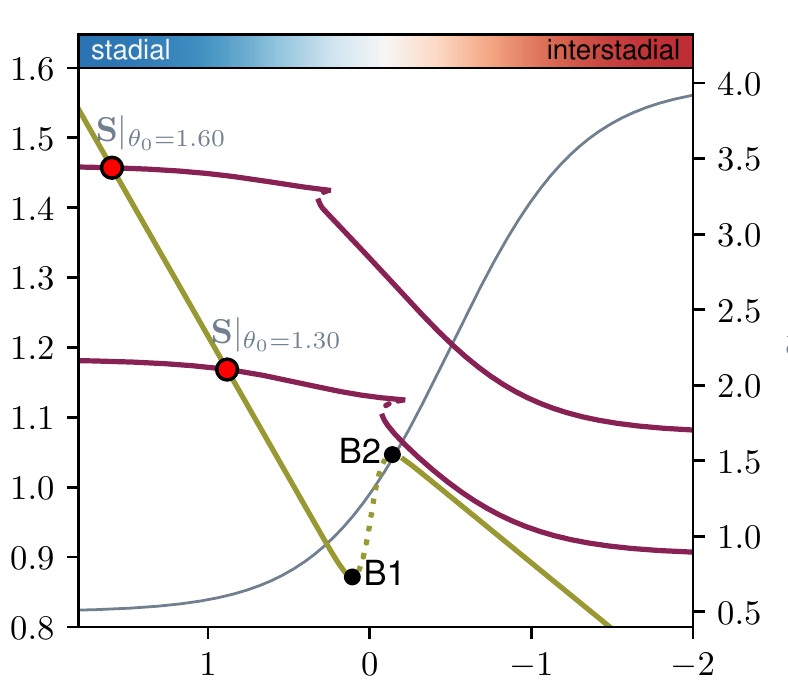}
  \caption{Nullcline of the seasonally averaged sea ice $I$ (olive)
    together with the nullcline of the Stommel atmosphere \(\theta\)
    (wine). 
    Due to the ice-albedo feedback the sea ice model features a bistable
    region where an ice-rich and a low-ice solution coexist. The difference
    in the slope of the two stable branches is controlled by the strength
    of the sea ice export \(R_{0}\). The lower \(\theta\)-nullcline is the
    same as in Fig.~\ref{fig:gamma_bifurcation} with \(\theta_{0} = 1.3\)
    upon using the transformation $\gamma=\gamma(I)$ given by
    Eq.~\ref{eq:A5}. The upper \(\theta\)-nullcline 
    results from setting \(\theta_{0} = 1.6\). The mutual relaxation rate
    \(\gamma(I)\) as a function of the sea ice is shown in light gray on
    the right ordinate. Intersections of the \(\theta\) and \(I\)
    nullclines define fixed points of the entire coupled system defined by
    Eqs.~\ref{eq:xStommel_short} and \ref{eq:A4}. Larger atmospheric
    backgrounds \(\theta_{0}\) yield more pronounced stadial conditions
    with colder Arctic atmosphere and larger sea ice cover.
  }
  \label{fig:sea_ice_bifurcation}
\end{figure}

As previously mentioned, sea ice insulates the atmosphere and the ocean from one
another and hence controls their mutual heat flux.
Based on our altered interpretation of the sea ice
variable \(I\), we model the effect of the sea ice on the ocean--atmosphere
heat flux in form of a hyperbolic tangent with saturation towards high
and low values of sea ice
\begin{equation}
\label{eq:A5}
\gamma(I) = \gamma_{0} +  \frac{\Delta \gamma}{2}\left[\tanh\left(\frac{-(I-I_{0})}{\omega}\right)+1\right].
\end{equation}
This yields a heat flux $\phi_{\mathrm{p}} \propto \gamma(I)\;(T-\theta)$ that in turn
controls the mutual atmosphere--ocean relaxation in terms of temperatures.
The parameters \(\gamma_{0}\) and \(\gamma_{0} + \Delta \gamma\) in
Eq.~\ref{eq:A5} define the mutual relaxation rate at maximum ice cover and
open ocean conditions, respectively, and \(\omega\) moderates the steepness
of the rate's decline with increasing sea ice thickness $I$.
Shifting the hyperbolic tangent along the sea ice axis by setting
\(I_{0} = -0.5\) yields an already substantially reduced heat flux
at intermediate sea ice cover (\(I=0\); cf.
Fig.~\ref{fig:sea_ice_bifurcation}). At maximum sea ice cover the heat
exchange between the high-latitude ocean and atmosphere is considered to be
(almost) shut off. At low latitudes, the heat flux is unaffected by the sea
ice. As explained in the derivation of the extended Stommel model, the term
\(\gamma(I)(T-\theta)\) is dominated by the heat flux between the polar
ocean and atmosphere as we have assumed
\(\theta_{\mathrm{e}}\sim T_{\mathrm{e}}\). Thus we choose a relatively
small \(\gamma_{0} = 0.5\) which reflects a substantially reduced total
atmosphere--ocean heat flux at maximum sea ice cover. The parameters for
the sea ice model were adopted from \cite{Eisenman2012} and
\cite{Lohmann2021a}. Only the newly introduced \(L_{0}\) and \(L_{1}\)
were tuned by hand to align the model's DO cycle shape with that observed
in the NGRIP record.


\paragraph{Dynamics of the coupled atmosphere--ocean--sea ice model.}
We now examine the joint bifurcation structure of the sea ice model and the
atmospheric temperature $\theta$ in the deterministic setting --- i.e. we
assess the nullclines of Eqs.~\ref{eq:xStommel_short} and \ref{eq:A4} (see
the main text for our working definition of the \textit{nullclines}). Our
sea ice model features a double fold bifurcation with respect to $\theta$
(Fig.~\ref{fig:sea_ice_bifurcation}). The associated bistability is an
effect of the ice--albedo feedback. The bifurcation points are given by
\(\mathrm{B}1 = (\theta = 0.83, I=0.11)\) and
\(\mathrm{B}2 = (\theta = 1.06, I = -0.14)\).
%
Intersections of the sea ice's and the atmosphere's nullclines constitute
fixed points of the full coupled system. The stable fixed points that
correspond to the two climate background states $\theta_{0}=1.3$ and
$\theta_0=1.6$ are
\((\theta_{\mathrm{s}} \approx 1.17, I_{\mathrm{s}}\sim 0.88)\) and
\((\theta_{\mathrm{s}} \approx 1.45, I_{\mathrm{s}}\sim 1.59)\),
respectively, and represent stadial climate states with large sea ice cover
and cold temperatures over Greenland (red dots in
Figure~\ref{fig:sea_ice_bifurcation}). Note that larger atmospheric
background gradients yield a more severe stadial climate with larger
\(I_{\mathrm{s}}\) and \(\theta_{\mathrm{s}}\). The corresponding stadial
fixed point in the ocean model, i.e. in Eqs.~\ref{eq:xStommel_short}, are
\(T_{\mathrm{s}} \approx 0.3\) and \(T_{\mathrm{s}} \approx 0.36\),
implying relatively warm Nordic Seas. Moreover, the AMOC strength
(\(q_{\mathrm{s}} \sim - 0.1\) and \(q_{\mathrm{s}} \sim - 0.08\)) is
negative and thus salinity driven and weak.

A key feature of the bifurcation diagram is the proximity of the low-ice
stable branch of the sea ice nullcline and the strong-mode stable branch of
the \(\theta\) nullcline around $I\approx -0.4$. The closer the nullclines
are to each other, the slower is the deterministic dynamics in nearby
regions of the state space. Their distance is controlled by the atmospheric
climate background \(\theta_{0}\) and decreases with decreasing $\theta_0$
provided that \(\theta_0>1.275\). If the nullclines are sufficiently close,
once the system enters this region of the state space, the dynamics allows
for prolonged periods in which $I$ and $\theta$ do not vary much, giving
rise to what we term a meta-stable state. It is the existence of this
transient meta-stable state that allows us to model prolonged
interstadials typical for DO cycles.


To bring the system close to the meta-stable state we require sufficiently
large perturbations. In our model these are provided by substantial
stochastic sea ice removals which reactivate the ocean--atmosphere
interaction and thereby trigger temporary state space excursions into the
interstadial regime. This is achieved by introducing a non-Gaussian
intermittent stochastic process which acts as a forcing on the sea ice and
is capable of inducing the required large abrupt sea ice removal.

\clearpage

\bibliographystyle{unsrt}  
\bibliography{bib}